\documentclass{article}
\pdfoutput=1

\usepackage{jheppub_mod}
\usepackage{cancel,multirow}
\usepackage[nameinlink]{cleveref}

\allowdisplaybreaks

\title{ALP-Pions generalized}
\author[a]{Triparno Bandyopadhyay,}
\author[b]{Subhajit Ghosh,}
\author[a]{and Tuhin S. Roy}
\affiliation[a]{
               Department of Theoretical Physics, 
               Tata Institute of Fundamental Research,
               Mumbai 400005,  India.
           }
\affiliation[b]{ 
               Department of Physics, 
               University of Notre Dame,
               South Bend, IN 46556, USA.
           }
\emailAdd{triparno@theory.tifr.res.in}
\emailAdd{sghosh5@nd.edu}
\emailAdd{tuhin@theory.tifr.res.in}

\preprint{TIFR/TH/22-35}
\abstract{
    A light axion-like particle or an ALP not just gives rise to
    interesting and spectacular signals of new physics as final states
    in meson decays, it necessarily leaves tell-tale signatures in
    processes that involve standard model (SM) fields only (i.e., SM
    processes). These effects result in the violation of the
    Gell-Mann--Okubo mass relation, modified form factors, altered
    integrated and differential rates for various SM transitions
    etc. This suggests that in the presence of a low lying state, such
    as an ALP, extraction of masses, mixing angles, and form factors in
    an entirely data-driven way from meson-physics observables is a
    highly non-trivial exercise.  However, once done correctly, these
    same observables may, in turn, provide important (indirect) bounds
    on ALP physics, which remain robust even in the limits where new
    physics effects conspire to weaken the bounds from direct searches.
    Starting with a generalized ALP-quark Lagrangian (where restrictions
    due to parity are removed) we demonstrate this approach by focussing
    on \texorpdfstring{\(K^+_{\ell_3}\)}{Kl3} decays, where we derive
    (indirect) bounds on ALP physics using NA48/2 data and lattice
    results. We also find sum rules which not just show deviations in
    the presence of an ALP, but also give hints towards the specific
    nature of the ALP physics itself.
}

\begin{document}
\maketitle

\section{
    Introduction
    \label{sec:intro}
}

Studied primarily in the context of the strong-CP problem
\cite{tHooft:1976rip, Peccei:1977hh, Peccei:1977ur, Weinberg:1977ma,
Wilczek:1977pj} and cold dark matter \cite{Preskill:1982cy,
Abbott:1982af, Dine:1982ah}, sub-GeV pseudo-Nambu-Goldstone bosons
(pNGB) are generic predictions of various new physics (NP) scenarios.
These range from the dynamical generation of small neutrino masses
(Majorons) \cite{chikashige:1980ui}, models attempting to solve the
flavor problem (Flavons) \cite{Froggatt:1978nt}, to models of universal
extra dimensions \cite{Izawa:2002qk} and string compactification
\cite{Witten:1984dg, Svrcek:2006yi, Arvanitaki:2009fg}.  Such pNGBs are
also invoked in the context of the anomalous muon magnetic moment 
\cite{Chang:2000ii}, the hierarchy problem \cite{Graham:2015cka},
electroweak baryogenesis \cite{Jeong:2018jqe}, and as a portal to dark
matter \cite{Nomura:2008ru}.  These pNGBs are typically symmetric under
a continuous shift of the field or---in less stringent cases---a
periodic shift. For the prototypical axion \cite{Weinberg:1977ma,
Wilczek:1977pj}, QCD breaks the shift symmetry giving a mass to it at
the corresponding scale (\(\Lambda_\mathrm{QCD}\)).  Yet, in general,
one can decouple the mass of the pNGB from \(\Lambda_\mathrm{QCD}\)
\cite{Rubakov:1997vp, Hook:2014cda, Fukuda:2015ana,
Marques-Tavares:2018cwm, Gherghetta:2020keg}.  Such pNGBs of diverse
origins and interactions---the masses of which are not strictly tied to
\(\Lambda_\mathrm{QCD}\)---are generally clubbed together in the
literature under the hypernym of Axion-Like Particles (ALP).  In this
paper, we define an ALP, \(a\), to be a pNGB with a periodic symmetry (compact,
e.g., when the centre symmetry of a non-abelian group is
preserved) that couples to QCD through dim-5 operators involving the
quarks and the dim-5 \(a G \tilde{G}\) coupling. We assume the presence
of a shift-breaking spurion contributing to the mass of the ALP.  

Owing to the diverse origins of ALPs, the detection strategies are
varied, with different experiments---both dedicated and
multi-purpose---probing different regions of the parameter space
\cite{Irastorza:2018dyq, Irastorza:2021tdu, Alves:2017avw,
Marciano:2016yhf, Jaeckel:2015jla, Dobrich:2015jyk, Knapen:2016moh,
Bauer:2017ris, Bauer:2018uxu, CidVidal:2018blh, Aloni:2018vki,
Bauer:2021mvw, Chakraborty:2021wda, Bertholet:2021hjl}.  The standard
sources of constraints on ALP couplings to QCD are Meson decay
experiments \cite{Freytsis:2009ct, Aloni:2018vki, Bjorkeroth:2018dzu,
    Altmannshofer:2019yji, Ishida:2020oxl, Bertholet:2021hjl,
Chakraborty:2021wda, PIENU:2021clt, Bauer:2021mvw}. The bulk of the
analyses focuses on looking for ALPs in the decay products of heavier
mesons. The SM interactions of the \(\pi^0\) source these processes,
with the ALP being emitted in place of the \(\pi^0\)---due to
\(a\)-\(\pi^0\) mixing. Even though direct detection experiments are the
most straightforward ways to look for a particle, these
are somewhat impeded by assumptions about the properties of
the hypothetical particle. These assumptions are related to the
decay length (prompt/displaced/invisible), the decay channel, and the
mass of the particle.  The direct detection bounds are effective in
constraining the Wilson coefficients corresponding to the ALP
interactions with the mesons, insofar as allowed by these underlying
assumptions. We can translate the bounds on the IR  coefficients to
those at the Electroweak (EW) scale, given a proper treatment of the
corresponding effective field theories (EFT)---matched at (\(\sim)\;
\Lambda_\mathrm{QCD}\).  The pioneering work in this regard
was presented in their 1986 paper by Georgi et al. \cite{Georgi:1986df}.
Recently, over multiple papers, Bauer et al. expands
\cite{Bauer:2020jbp} on the original work and points out a key
omission \cite{Bauer:2021wjo}.  In the latter, the authors present a
meticulous matching of the chiral Lagrangian to the EW Lagrangian and
then show the effects of the new operator on ALP-meson amplitudes. 

We begin this work in a similar spirit and construct the chiral
perturbation theory in the presence of a light ALP (A\(\chi\)PT).  We
call this setup `generalized' in the sense that we derive the
A\(\chi\)PT from a larger set of operators (up to \(d=5\)) involving the
ALP and the quarks, where only the compactness criterion of the ALP is
taken into account. Extending the framework to include operators where
the ALP appears as a scalar and not a pseudo-scalar, allows us to
include theories where the ALP may mix with CP even states. More
importantly, it enables us to perform phenomenological studies of models
where light CP-even scalars interact with the SM through a Higgs-portal
like setup \cite{Patt:2006fw}.  We construct the ALP-quark lagrangian in
the EW basis where left-handed `u' and `d' quarks are treated on an
equal footing. One advantage of such a basis is that it helps us to
identify relationships among Wilson coefficients in the A\(\chi\)PT,
which are hard to guess otherwise, and also to identify Wilson
coefficients which have an additional suppression proportional to the EW
breaking vev. Note that, in this work, we refrain from working with
`all' possible dim-5 contact terms between the ALP and the quarks. 
 
Apart from working with a generalized version of the A\(\chi\)PT, this
work differs from the bulk of the literature on ALPs in that we focus
exclusively on processes that involve SM final states only. To be
specific, we look at how the charged current (CC) interactions of the
pions get modified in the presence of the low-lying ALP. The
implications of the existence of a light ALP are many, starting with the
modification of the meson mass spectra over the predictions of the SM
chiral lagrangian (SM\(\chi\)PT). A particularly striking effect of this
modification is the violation of the Gell-Mann--Okubo mass (GMO) mass
relations \cite{Gell-Mann:1962yej,Okubo1} at tree level.  We also find
important deviations of the form factors (FFs) in the A\(\chi\)PT, when
compared to the SM\(\chi\)PT \cite{Leutwyler:1993iq} . Take for example
the FFs \(f_\pm^{K\pi}(0)\) defined as the matrix element \(\langle \pi
| \bar{s}_L\gamma^\mu u_L|K\rangle\), where the \(\pm\) refers to the
components of the matrix element along \(p_K\pm p_\pi\) respectively.
We find that
\begin{align}
    \label{eq:introeq}
    \frac{f_+^{K^+\pi^0}(0)}{f_+^{K^0\pi^+}(0)}
    \Big|_{\mathrm{A}\chi\mathrm{PT}}=
    \frac{f_+^{K^+\pi^0}(0)}{f_+^{K^0\pi^+}(0)}
    \Big|_{\mathrm{SM}\chi\mathrm{PT}}+\frac{f_\pi^2}{f_a^2} K_1\;, 
\end{align}
where \(f_\pi\) and \(f_a\) are the characteristic scales associated
with the pions and the ALP respectively, and \(K_1\) is a function of
various Wilson coefficients of the ALP-quark contact operators. 

As expected, all these results simply suggest that even the rates of
various processes in the SM meson sector will invariably deviate in the
presence of an ALP. As a specific example, we consider in this work the
CC decay of the charged Kaon to a neutral pion and two leptons,
\(K^\pm\to \pi^0 \ell\nu\) (\(K^+_{\ell_3}\)). The decay
amplitudes computed in the A\(\chi\)PT deviate from that in the SM
primarily because of two reasons. Firstly, the matrix element of the SM
operator changes as neutral pions mix with the ALP (this effect is
captured in the altered FFs), and secondly, there exist new operators
which contribute at the order \((f_\pi/f_a)^2\). The decay width and the
corresponding differential distributions, therefore, differ from the SM
expectations at \(\mathcal{O}\left(\frac{f_\pi^2}{f_a^2}\right)\). This
leading order effect appears as interference between the SM piece and
the NP part of the amplitude.  

This presents us with a dilemma as well as an opportunity.  Clearly, the
extraction of SM parameters (masses, mixing angles, FFs etc.) in a
data-driven way from pion data becomes a nontrivial exercise in the
presence of a low-lying ALP. In fact, one requires a systematic study
where these quantities are either extracted from observables that remain
relatively unaffected and/or calculated theoretically (such as lattice).
Take for example, the CC interactions of mesons, which are standard
sources for extracting CKM elements. The rate of  \(K_{\ell_3}\), used
as a standalone measurement of \(V_{\bar{s}u}\), and the differential
distributions,  used for data-driven determination of the
\(f^{K\pi}_\pm\) FFs, are both sensitive to NP effects even at tree
level. On the other hand, if one uses \(V_{\bar{s}u}\) extracted from
channels unaffected by the ALP physics and FFs from lattice
measurements, one can turn the argument around and use these precision
measurements to constrain the ALP physics. 

As a concrete demonstration, we use the observed \(K^+_{\ell_3}\) 2D
Dalitz distribution as reported by the NA48/2 collaboration
\cite{Lazzeroni:2018glh} and the particle data group (PDG) average of
the partial \(K^+_{\ell_3}\) width \cite{Zyla:2020zbs} to constrain the
parameters of the A\(\chi\)PT. As SM inputs, we use \(V_{\bar{s}u}\)
extracted from \(K^+\to \mu^+\nu\) \cite{Zyla:2020zbs} and the lattice
computations of the \(f^{K\pi}_\pm\) as reported by the European Twisted
Mass Collaboration \cite{Carrasco:2016kpy}.  We then go on to provide
the first, to our knowledge, indirect constraints on the A\(\chi\)PT
parameters using \(K^+_{\ell_3}\). At first glimpse, the constraints
appear much weaker than those from direct ALP searches.  However, these
constraints are somewhat model independent --- in that these are largely
independent of the details of the ALP mass, decay channels, and
lifetime. More importantly, we show that these limits remain valid even
in the corners of the theory space where the \(K^+\) width to
\(a\ell\nu\) is suppressed. Note that, the choice of the 
CC decay of the \(K^\pm\) over the much simpler \(\pi^\pm\) system (say,
\(\pi_\beta\)) is motivated by the identification that there are
multiple observable effects of the ALP which are exhibited only in
\(K^\pm\) decays and not in those of the \(\pi^\pm\).

Finally, we construct  sum rules out of the FFs corresponding to the CC
semi-leptonic decays of the mesons.  The sums can indicate the presence
and the nature of an ALP in the chiral Lagrangian. For example,  the sum
corresponding to the \(K^+\) FFs is identically equal to one in the SM,
deviating from unity, at tree level, in the presence of the ALP.  What
is striking is that it can deviate on either side of unity, the sign of
the deviation pointing to qualitatively different kinds of UV physics. 

The paper is structured as follows. In the following section
(\cref{sec:lag}) we write down the dim-5 operators of the ALP in the EW
symmetric phase of the SM and go on to match those onto the chiral
Lagrangian. We go on to show the deviations to SM expectations in
presence of the ALP.  In \cref{sec:pheno} we focus on \(K^+_{\ell_3}\)
decays. We first list all the different ALP sources modifying the
\(K^+_{\ell_3}\) amplitude  and then go on to constrain these NP effects
using data on \(K^+_{\ell_3}\) decay spectra. In \cref{sec:ida}, we
discuss new contributions to the amplitudes with the ALP in it and
discuss limits where the ALP amplitude is subdominant compared to the
corresponding modification to SM amplitudes. In \cref{sec:sr}, we go on
to derive the sum rules discussed above. Finally, we conclude.

\section{
	Formalism: Construction of the general ALP-Pion Lagrangian
	\label{sec:lag}
}
Deriving the ALP-pion interactions in the IR is subtle, where subtleties
arise from the matching of the ALP-quark Lagrangian above the QCD scale
to the chiral Lagrangian. The most crucial piece in the calculation
stems from the choice of the basis in which the ALP-quark Lagrangian is
written. Before beginning the following subsection where we methodically
derive the ALP-pion Lagrangian, here we initiate a brief discussion
regarding the basis.

In order to substantiate the \emph{right} choice for the basis, first note
that non-trivial constraints exist in the IR Lagrangian, the origin of
which lies in the demand that the ALP-quark Lagrangian must arise from a
fully electroweak-symmetric theory at short distances.  This
observation gives the correct power counting (hence, the right
suppression) for specific terms in the ALP-pion chiral Lagrangian which
are harder to guess at low energy.  Consider, for example, the two seemingly
different operators, 
\begin{equation}
	\label{eq2.1}
	k_1 \frac{f_\pi}{2 f_a}  \:  \partial_\mu a\, \partial^\mu
	\pi_0  \qquad \mathrm{and}   \qquad
	i k_2 \frac{1}{2 f_a} \: \partial^\mu a
    \left(
        \pi^+ \partial_\mu\pi^{-} - \pi^- \partial_\mu\pi^{+}
    \right) \; .
\end{equation}
In the above, $k_1$ and $k_2$ are dimensionless coupling constants, and the
scales $f_a$ and $f_\pi$ are characteristic scales associated with the
ALP and the pion physics respectively. The first
operator generates the kinetic mixing between the ALP and $\pi^0$, while 
both these operators play important parts in pion decay. It turns out
that in the EW limit one finds  $k_1=  - k_2$. Consequently,
\begin{equation}
	\label{eq:eq2.2}
    k_1 + k_2 \simeq \mathcal{O}(v/\Lambda_\mathrm{UV}) \; ,
\end{equation}
if we assume that even in the presence of NP, the Higgs vev, $v$,
remains the only source of electroweak symmetry breaking. Note,
$\Lambda_\mathrm{UV}$ represents a high scale far above the EW scale,
for example, corresponding to a higher dimensional Higgsed operator.

Even though the statement in \cref{eq:eq2.2} appear to be highly
nontrivial, it can be easily understood when the low energy chiral
Lagrangian is derived from the manifestly EW symmetric  quark
Lagrangian.  In order to demonstrate it, we begin with QCD with the
number of flavors $N_f=2$.  The theory contains an approximate
$SU(2)_L\times SU(2)_R$, realized as
\begin{equation}
	\label{eq2.3}
	\text{using}  \qquad  q\equiv \begin{pmatrix} u \\ d \end{pmatrix} \; ,  \qquad
	q_{L} \equiv P_{L}  \: q  \xrightarrow[L\times R]{}  L\: q_L  \qquad \text{and} \qquad
	q_{R} \equiv P_R \:  q  \xrightarrow[L\times R]{} R \: q_R,
\end{equation}
where $P_L$ and $P_R$ are the projection operators that projects out the
left- and right-handed spinors of the Dirac fermions $u$ and $d$. The
contact operators relevant for our discussion are given as
\begin{equation}
	\label{eq:eq2.4}
	\sum_{i=0}^3 \ k^L_i  \: \frac{\partial_\mu a}{f_a} \overline{q}_L t^i \gamma^\mu q_L
	\; ,\qquad  \text{and}  \qquad
	\sum_{i=0}^3 \ k^R_i \:  \frac{\partial_\mu a}{f_a} \overline{q}_R t^i \gamma^\mu q_R   \; ,
\end{equation}
where $t^i$ are the generators of $SU(2)$ and $k^{L/R}$ play the role of
Wilson coefficients. Note that all such terms
are not allowed by the symmetries of the SM. The operators proportional
to $t^1, t^2$, or rather $t^{\pm}$, break  $U(1)$ electromagnetism and
are therefore not allowed. More constraints follow once we recognize
that  the $SU(2)_L$ in this particular case can be identified with the
electroweak $SU(2)_W$, and therefore $q_L$ transforms as a doublet under
$SU(2)_W$. Therefore, in the EW limit one obtains the result that all
$k^{L}_i\rightarrow 0$, except for $i =0$. Summarizing:
\begin{equation}
	\label{eq2.5}
	\begin{split}
		k_{1,2}^{L/R} \ = \ 0 \quad & : \quad \text{from Electromagnetic
        invariance},  \\
		k_{1,2,3}^{L} \ \rightarrow  \ 0  \quad & : \quad \text{in the EW symmetric limit}  \; .
	\end{split}
\end{equation}
We now show that \cref{eq:eq2.2} follows from the fact that $k_{3}^{L} \rightarrow 0 $ in the EW limit.

In order to derive the IR effective operators, which match to the
operators in \cref{eq:eq2.4} at the leading order, we use the
symmetry properties of the pion fields and current matching. To be
specific,  note the symmetry transformation of the pion field and
the corresponding currents under $SU(2)_L\times SU(2)_R$:
\begin{equation}
	\label{eq2.6}
	\begin{gathered}
		U_\pi \ \equiv \ \exp\left( \frac{2i\pi^a t^a}{f_\pi}  \right)   \quad \xrightarrow[L\times R]{}  \quad
		L \: U_\pi \: R^\dag  \\
        \implies 	J^a_{L_\mu}	 \ = \  - i \frac{f_\pi^2}{2}
        \mathrm{Tr}\!\left[U_\pi^\dagger t^a \partial^\mu U_\pi\right]+\cdots
		\; \text{whereas} \;
		J^a_{R_\mu}	 \ = \  - i \frac{f_\pi^2}{2}
        \mathrm{Tr}\!\left[U_\pi  t^a \partial^\mu
        U_\pi^\dag\right]+\cdots \;.
	\end{gathered}
\end{equation}
Therefore, at leading order, the interactions between the axion
and the pions can be derived by simply matching currents. In
particular, consider the operators with Wilson coefficient $k_3^L$ and
$k_3^R$ in  \cref{eq:eq2.4}:
\begin{align}
	k^L_3 \:  \frac{\partial_\mu a}{f_a} \overline{q}_L t^3 \gamma^\mu q_L  \quad
	& \rightarrow \quad  k^L_3 \:  \frac{\partial_\mu a}{f_a}  J^3_{L_\mu}
	+\cdots \; \rightarrow	- \ i k^L_3  \: \frac{f_\pi^2}{2 f_a} \:
    \partial_\mu a \:  \mathrm{Tr}\left[U_\pi^\dag t^3\partial^\mu
    U_\pi\right] +\cdots \;,
	\label{eq2.7} \\
	k^R_3 \:  \frac{\partial_\mu a}{f_a} \overline{q}_R t^3 \gamma^\mu q_R  \quad
	& \rightarrow \quad  k^R_3 \:  \frac{\partial_\mu a}{f_a}  J^3_{R_\mu}
	+\cdots\; \rightarrow
    - \ i k^R_3  \: \frac{f_\pi^2}{2 f_a} \:  \partial_\mu a \:
    \mathrm{Tr}\!\left[U_\pi t^3\partial^\mu U_\pi^\dag\right] +\cdots
    \;.
	\label{eq2.8}
\end{align}
Expanding the exponential, one finds the interactions listed in
\cref{eq:eq2.4}, with the identification
\begin{equation}
	\label{eq2.9}
	k_1 \ = \ k_3^L - k_3^R  \qquad \text{and} \qquad k_2 \ = \   k_3^L
    + k_3^R \;\;.
\end{equation}
The conditions referred to in \cref{eq:eq2.2} therefore follows from
noting that $k_3^L=\mathcal{O}(v/\Lambda_\mathrm{UV}) $.

The discussion above should make it clear that it is important to choose
the \emph{right} basis in order  to write the general ALP-quark
Lagrangian before matching, where the right basis is where the EW
symmetry (as a global symmetry) in the ALP-quark interactions is
manifest and is only broken by the EW-breaking parameters of the SM.
However, this consideration does not uniquely define the basis, and
additional redefinitions ($a$-dependent) of the quark fields (both chiral
and vector in nature) are allowed that keep the EW symmetry manifest. A
convenient use of the chiral redefinition can be invoked to get rid of
the operator $ aG\widetilde{G}$. In this particular case, the chiral
rotation must be flavor universal. Therefore, we pick the right basis to
be the basis with manifest EW symmetry and without the $
aG\widetilde{G}$ operator.  Further, any new flavor universal vectorial
redefinition of the quark fields does not introduce any new independent 
operator.

\begin{figure}[htpb]
	\centering
    \includegraphics[scale=1]{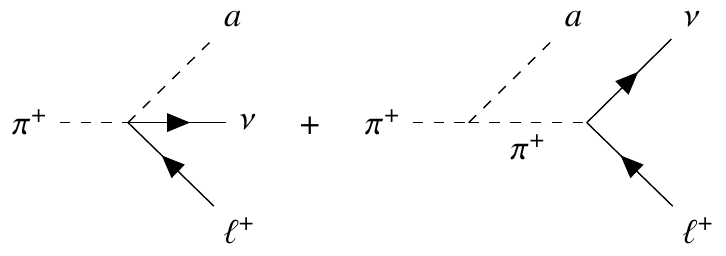}
    \caption{Diagrams contributing to \(\pi^+\to a \ell^+\nu\) in a
    generic basis.}
	\label{fig:diag}
\end{figure}

To expand, it is instructive to compare the fate
of \cref{eq:eq2.2}, in the `wrong' basis.  It should be clear from the
discussion above that a basis where the components of the electroweak
doublet (namely, $q_L$) are treated differently can be obtained from the
`right' basis after a flavor dependent redefinition of the quark fields
is performed.  However, since electromagnetism remains a good
symmetry in the IR, we are left with only redefinitions generated by
$t^3$. Further, the redefinition must be vectorial to ensure that  the $
aG\widetilde{G}$ operator remains absent in the new basis.  It is
easy to check that such a redefinition keeps  $k_1$ the same but changes
$k_2$ and consequently \cref{eq:eq2.2}. Therefore, \cref{eq:eq2.2}
is true only  in the manifest EW basis.  Of course, there is no need to
assert that observable are basis independent. To confirm, 
consider the decay $\pi^+\to a \mu^+ \nu $.  Starting with the manifest
EW basis, we perform a flavor dependent vectorial redefinition
(parameterized by, say, $\zeta$) of the quark fields:
\begin{equation}
	\label{eq:eq2.10}
	\begin{gathered}
		q \rightarrow \exp\left( i \zeta \frac{ a }{f_a}  t^3 \right) q;  \\
        \mathcal{L} \rightarrow \mathcal{L} +
		\zeta\frac{\partial^\mu a}{f_a}
        \left(\pi^- \partial_\mu\pi^{+} - \pi^+ \partial_\mu\pi^{-}\right)
		+  2 G_F \zeta \frac{f_\pi}{f_a}  \left( \partial_\mu a \pi^{+}
		+ a \partial_\mu \pi^{+} \right)
        \overline{\ell}\gamma_\mu  P_L \nu  + \cdots \; .
	\end{gathered}
\end{equation}
In the above, \(\cdots\) refer to other deviations which do not impact the
decay  $\pi^+\to a \mu^+ \nu$ at the leading order. The new
contributions to the amplitude proportional to $\zeta$ is shown in
\Cref{fig:diag}. It is easy to check that the contributions to the
amplitude get neatly canceled out between the two diagrams. As expected,
there is no effect in the amplitude for observables even though the
redefinition in \cref{eq:eq2.10} gives rise to a  large number of
operators  (proportional to $\zeta$). In the next subsection, when we
work with three flavors, we will see that the three-scalar interaction
is relevant even if we work in the manifestly EW symmetric basis.
Indeed, for the three flavor case, the left-handed interactions
corresponding to \(t^8\) of \(SU(3)\) are EW symmetric while being
flavor dependent. Therefore, there will be a three-scalar interaction
present even in the manifestly EW basis.

\subsection{
    The chiral Lagrangian with the ALP: \texorpdfstring{A\(\chi\)PT}{AkPT}
    \label{sec:chialp}
}

In this subsection, we begin with and categorize the general ALP-quark
Lagrangian in the manifestly EW basis, where EW symmetry is realized as
a global symmetry and is only broken by the Higgs vev.  As described
before, this basis is also characterized by `no' \(a G\widetilde{G}\)
operator.

Before proceeding, however, note that we discuss now an \(N_f = 3\)
scenario, which is characterized by an approximate global \(U(3)_L\times
U(3)_R(\equiv SU(3)_L\times SU(3)_R\times U(1)_L\times U(1)_R)\). The EW
factor \(SU(2)_W\) is identified as an \(SU(2)\) subgroup of \(U(3)_L\),
\(SU(2)_L\), which suggests that EW invariant but flavor dependent
redefinitions of the quark triplets exist. In particular, by choosing the
quark basis as in the equation below, the \(SU(2)_L\)  can be understood
to have been embedded in the top-left \(2\times 2\) block of the
\(SU(3)_L\).  Under the action of \(SU(3)_L\times SU(3)_R\), we have,
\begin{equation}
	\label{eq3.1}
	\mathrm{using}\quad  q\equiv \begin{pmatrix} u \\ d \\s \end{pmatrix},
    \quad q_{L} \equiv P_{L}  \: q  \xrightarrow[L\times R]{}  L_3\: q_L
    \quad \text{and} \quad
	q_{R} \equiv P_R \:  q  \xrightarrow[L\times R]{} R_3 \: q_R \; ,
\end{equation}
where $L_3$ and $R_3$ represent transformation matrices for the
$SU(3)_L$ and $SU(3)_R$ respectively. In this basis, the structure of
the EM charge matrix is
\begin{equation}
	\label{eq3.2}
	Q\ = \ \begin{pmatrix}
        2/3 & 0 &  0 \\
        0 & -1/3 & 0 \\
        0 & 0 & -1/3
    \end{pmatrix}\ = \  t^3 + \frac{1}{\sqrt{3}} t^8\;.
\end{equation}
Finally, note that as far as the ALP is considered, we only impose the
compactness condition,
\begin{align}
    \label{eq:AlSymm}
    a \to a+2n\pi f_a,\; n\in \mathbb{Z}.
\end{align}
Operators involving derivatives of the ALP automatically satisfy the
compactness criterion. In fact, these are additionally symmetric under
any shift of the ALP (\(n\in \mathbb{R}\))---a property of the NGBs.  In
our generalized ALP-Quark framework we stick to the compactness
condition, and do not extend it to the full shift symmetry of an NGB.
This  allows for the presence of sinusoidal functions of \(a\). Note
that, the polynomial $\sin na/f_a$ is especially important since it can
provide a term linear in \(a/f_a\).

With the symmetry properties of the ALP defined, we are ready to
categorize the interactions of the ALP with the SM quarks at the leading
order (linear in \(a/f_a\)). In particular, we find that all such
interactions can be broadly categorized in five classes of operators
involving the ALP and various quark bilinears. These operators are:
\begin{equation}
	\label{eq:eq3.3}
	\sum_{i=0}^8 \left( 
        C_{L}^i \mathcal{O}_{L}^i + C_{R}^i \mathcal{O}_{R}^i   
        + C_{LR}^i \mathcal{O}_{LR}^i  
    \right) 
        + C_{W} \mathcal{O}_{W}   
        + C_{Z} \mathcal{O}_{Z} 
      \; ,
\end{equation}
where $C_a$s are \emph{real} Wilson coefficients for the operators
belonging to the class $\mathcal{O}_a$.

\begin{table}[htpb]
	\centering
	\begin{tabular}{|c|c|l|}
    \hline
    &&\\
    \multirow{4}{*}{$\mathcal{O}_{L}^i$}
        &\multirow{4}{*}{
               $\dfrac{1}{f_a} \partial_\mu a\:
                \overline{q}_L t^i \gamma^\mu q_L$
            }
        &$i = 0,8:$ Allowed \\
        && $i = 1,2,4,5:$ break EM \\
        &&$i = 1(1)7:$ break $SU(2)_W$\\
        && $i = 6,7:$ tree FCNC \\
        &&\\
    \hline
    &&\\
    \multirow{2}{*}{$\mathcal{O}_{R}^i$}
        &\multirow{2}{*}{
            $\dfrac{1}{f_a}\partial_\mu a\: \overline{q}_R t^i\gamma^\mu q_R$
            }
        &
            \\
    &&\(i=0,3,8:\) Allowed\\
    &&\\
    \cline{1-2}
    &&$i = 1, 2,4,5:$ break EM\\
    \multirow{2}{*}{ $\mathcal{O}_{LR}^i $ }
        &\multirow{2}{*}{
            $\dfrac{a}{f_a}  \:  \overline{q}_L t^i  M q_R$
        }
        &$i = 6,7:$ tree FCNC\\
        &&\\
        &&\\
        \hline
        &&\\
     $\mathcal{O}_{W} $
        & $-\dfrac{a}{f_a}  \: \overline{q}_L 
			Q^W \cancel{j}_\pm q_L$
            
        &\\
        && \\
        \cline{1-2}
        && \\
    $\mathcal{O}_{Z} $ 
        &
    $-\dfrac{a}{f_a}  \:\left( \overline{q}_L
			         Q_L^Z\; \cancel{j}_Z q_L
                 + \overline{q}_R Q_R^Z\; \cancel{j}_Z q_R \right)$
        &\\
    &&\\
    \hline
	\end{tabular}
    \caption{We tabulate the  dimension-5 operators generated by the
        presence of an ALP in the EW basis. We indicate the generators
        allowed by symmetry for each of the operators and indicate the
        ones we have dropped.
    }
	\label{tab:1}
\end{table}

We tabulate the operators in each of these categories in \Cref{tab:1},
where $t^0$ is the identity matrix and $t^{1(1)8}$ are generators of
$SU(3)$. The \(j^\mu_\pm\) current consists of quark and lepton
bilinears, which arise as we replace the EW field \(g_W W^\mu_\pm\to 4
G_F j^\mu_\pm\)  using the equation of motion in order the match to the
EFT (Fermi theory) at the  EW scale.  Similarly, \(j_Z^\mu\) is the
replacement for the EW gauge boson \(Z^\mu\). The matrices \(Q^{W}\) and
\(Q^Z_{L/R}\) are the three-dimensional representations of the strengths, as
in the SM, corresponding to the \(j^\mu_\pm\) and \(j_Z^\mu\) currents
respectively.  We match the operators at the EW scale at the tree level
and do not take into consideration effects due to renormalization below
the EW scale.  The strength of these interactions is of course given by
the Fermi constant, defined as \(G_F=\sqrt{2}g_W^2/(8M_W^2)\).  Before
proceeding, first note the important points regarding \Cref{tab:1}:
\begin{itemize}
    \item Operators of the type  $\mathcal{O}_{LR}^i $,
        $\mathcal{O}_{L}^i $, and  $\mathcal{O}_{R}^i $ have been
        studied exhaustively
        \cite{Georgi:1986df,Aloni:2018vki,Bauer:2021wjo,Cheng:2021kjg}. The first type
        ends up giving rise to the mass mixing between the ALP and
        pions, whereas operators of the second and the third kind
        generate kinetic mixings as discussed in the last subsection.
    \item Additional comments are due regarding the operators
        \(\mathcal{O}_{L}^0\) and  \(\mathcal{O}_{R}^0\), which
        correspond to ALP derivative couplings to the  $U(1)_L$ and $
        U(1)_R$ currents respectively, out of which the axial
        $U(1)_{L-R}$ is anomalous and broken by QCD. Not surprisingly,
        one finds that a nonzero  $\left(C_{L}^0 - C_{R}^0\right)$
        generates kinetic mixing between the ALP and the $\eta^\prime$
        meson. Since we are primarily interested in the phenomenology of
        the pion octet in the presence of the ALP, we set $C_{L}^0 =
        C_{R}^0 = 0$ for subsequent discussions and leave the  exercise
        of including the \(\eta^\prime\) meson for future endeavours.
    \item All operators listed in \cref{eq:eq3.3} are not allowed.  The
        third column in \Cref{tab:1} lists operators which violate EM
        and therefore cannot be present in the IR. Similarly, we also
        mention the operators which violate the global $SU(2)_W$. As
        argued before, these must be suppressed by additional powers of
        EW breaking parameters. Finally, note that the operators need
        to be rewritten in the quark mass basis before we can derive the
        A\(\chi\)PT.  This introduces new tree level FCNC operators. For
        simplicity, we assume small \(C^6\) and \(C^7\) in the EW basis
        which ensures (after appropriate tuning) that there in no FCNC
        in the ALP-quark EFT in the mass basis. For a detailed
        discussion of flavor violation generated by the ALP-quark
        Lagrangian check Ref.  \cite{Bauer:2021mvw}.
\end{itemize}

As mentioned before, operators of the type  $\mathcal{O}_{LR}^i $,
$\mathcal{O}_{L}^i $, and  $\mathcal{O}_{R}^i $ are well studied. On the
other hand, operators $\mathcal{O}_{W}$ and $\mathcal{O}_{ZL/R}$,  are
not. In fact, we are not aware of any previous literature which includes
these operators in the derivation of the ALP-meson EFT. There are many
ways via which these operators can be generated. Here we give one
possible UV scenario which results in an EFT with $\mathcal{O}_W$ and
$\mathcal{O}_Z$. Begin with a fully EW (gauged) symmetric theory and
consider the following potential involving the Higgs doublet:
\begin{equation}
	V\left(H, a \right)=  
        - \mu^2\!\left(a\right)H^\dag H 
        + \frac{1}{2}\lambda\!\left(a\right) \left( H^\dag H \right)^2\;.
\end{equation}
In the above, we have simply extended the SM parameters in the Higgs
potential to include polynomials of sinusoidal functions of the ALP,
making the compactness condition manifest. The simplest way to
analyze this theory is to expand it around the $a$-dependent minimum,
bringing in $a$-dependent masses for the EW gauge bosons. Keeping
terms linear in $a$, we find that a replacement of the EW vev (and
therefore of the Fermi constant) by its $a$-dependent value allows us to
derive the low energy EFT:
\begin{equation}
	\begin{aligned}
        v^2\rightarrow v^2\!\left(a\right)= 
            \frac{\mu^2\!\left(a\right)}{\lambda\!\left(a\right)} 
            \equiv v^2 \left(1 - C_v \frac{a}{f_a} + \dots \right)&\\
        \implies   G_F  \rightarrow  G_F\!\left(a\right) 
        =  G_F \left( 1 + C_v \frac{a}{f_a} + \dots \right)&.
	\end{aligned}
\end{equation}
One finds $\mathcal{O}_W$ and $\mathcal{O}_Z$ with the
constraint that $C_W = C_Z =   C_v$. 

Coupling the ALP to the scalar potential allows us to incorporate
Higgs-portal scenarios into our EFT.  In such models, the hidden sector
talks to the SM through the only available super-renormalizable term in
the SM, the Higgs quadratic term \cite{Patt:2006fw}. This allows us to
include the phenomenology of light scalars in pion decays even for the
cases where the light scalar is not in an exact parity eigenstate. 

Note that, even though we talked about the \(aG\tilde{G}\) operator (and
subsequently chose a basis where it is eliminated), we do not discuss
other contact operators between the ALP and gauge field bilinears. We
note, the operators \(aW\tilde{W}\) and \(aB\tilde{B}\)---\(W\) and
\(B\) being the field strengths corresponding to \(SU(2)_W\) and
\(U(1)_Y\) respectively---can be eliminated by redefining the lepton
doublets and singlets. Of the other operators, namely, \(aG^2\),
\(aW^2\), and \(aB^2\), the effects of the latter two are already
captured by \(\mathcal{O}_{W/Z}\). This is because, one can incorporate the
effects of these operators by simply substituting the EW gauge couplings
\(g_{W/Z}\to g_{W/Z}(1+ C a/f_a)\) which in turn generate \(\mathcal{O}_{W/Z}\).
As for the \(aG^2\) operator, we ignore it here, and will follow up on
it in a future work.  However, do note that the \(aG^2\) operator can
only appear at one loop order and therefore is small.

In order to derive the ALP-pion Lagrangian, we write the quark
Lagrangian in a convenient form,
\begin{align}
\label{eq:eq2.14}
    \mathcal{L}\ \supset\  &
        \overline{q}_L \gamma^\mu
        \left(i \partial_\mu +   L_\mu \right)  q_L
        + \overline{q}_R\gamma^\mu
        \left(i\partial_\mu + R_\mu \right) q_R
        + \overline{q}_L  \overline{M} q_R
		+ \cdots \;\text{, where} \notag \\
    &L^\mu \ = \ 
		Q A^\mu
        +
        \left(1+C_Z \frac{a}{f_a}  \right)
		Q_L^Z j_Z^\mu        +
        \left(1+C_W \frac{a}{f_a}  \right)
		Q^W j_\pm^\mu
		+  \frac{\partial^\mu a}{f_a} C^8_{L} t_8  \; ,\notag\\
    &R^\mu \ = \ 
		Q A^\mu
        +
        \left(1+  C_Z \frac{a}{f_a} \right)
		 Q_R^Z j_Z^\mu
		+\frac{\partial_\mu a}{f_a} \sum_{i=3,8} C^i_{R} t^i\;,\notag\\
    &\overline{M}\ = \ \sum_{i}^{0,3, 8}\left(
            1+ i C^i_{LR} \frac{a}{f_a} t^i
            +\cdots
        \right)  M.
\end{align}
Here, we lightly gauge $SU(3)_L \times SU(3)_R$, with $ L_\mu,  R_\mu$
compensating for the corresponding gauge transformations of the quark
fields. With the $\overline{M}$ as a spurion, even the mass term acts
gauge invariant.

With the quark Lagrangian defined in the presence of the ALP, the pion
Lagrangian is obtained by current matching. Consider the fundamental
construct in the pion Lagrangian, $U_\pi$ (the exponential
representation of the pions \cite{Callan:1969sn}), which  transforms as
a bi-fundamental  under   $SU(3)_L \times SU(3)_R$:
\begin{equation}
	\label{eq:eq2.15}
	U_\pi \ \equiv \ \exp\left( \frac{2i\pi^a t^a}{f_\pi}  \right)   \quad \xrightarrow[L\times R]{}  \quad
	L_3 \: U_\pi \: R_3^\dag,
\end{equation}
so that, the `gauge-invariant' kinetic piece of $U_\pi$ and the mass
term nicely match to
\cref{eq:eq2.14}:
\begin{equation}
	\label{eq:eq2.16}
	\mathcal{L} \ \supset \   \frac{f_\pi^2}{4} \:  \text{Tr}
	\Big[\left| \partial_\mu U_\pi - i (L_\mu U_\pi - U_\pi R_\mu) \right|^2 \Big]  \ + \
	\frac{\Lambda f_\pi^2}{2} \: \text{Tr} \Big[  \overline{M}   U^\dagger_\pi  \Big]  \ + \ \text{h.c.}
	\ + \ \cdots .
\end{equation}
Here, $\Lambda$ is the UV cut off of the EFT, and $\cdots$ represent
higher order terms in the chiral Lagrangian. The Lagrangian in
\cref{eq:eq2.16}, with \(L_\mu\), \(R_\mu\), and \(M\) given by
\cref{eq:eq2.14}, gives the leading order terms in the  A\(\chi\)PT.
This completely reduces to the familiar chiral Lagrangian in the limit
$f_a\rightarrow \infty$. 

While the introduction of the ALP fields by symmetry matching is straightforward, a subtle point needs 
to be addressed with regards to power counting.  
Before the addition of
external currents, there are two power counting parameters in the
chiral Lagrangian. The first of these is the momentum \(p^2/\Lambda^2\), where \(\Lambda\) is the
cutoff and is taken to be of the order of the \(\rho\) meson mass, the second is
the quark mass(es), \(m_q/\Lambda\) (see, e.g, \cite{Kaplan:2005es}). The external currents come
with their own power counting parameters, \(\alpha_{EM}\) for EM and
\(p^2 G_F\) for the electroweak currents. The ALP field, however,
   introduces additional derivatives through the external currents. The
   introduction of these `new' derivatives do not spoil the momentum power
   counting of the chiral Lagrangian. This is because the ALP derivatives come suppressed by 
   a characteristic scale, \(f_a\), of its own. In essence, there are two new power counting parameters that the ALP brings in, \(f_\pi/f_a\) and \(p_\mu/f_a\). Therefore, corresponding to each derivative of the `pure' chiral Lagrangian, the ALP dericvatives are \(f_\pi/f_a\) suppressed.

We break down the A\(\chi\)PT Lagrangian in \cref{eq:eq2.15}, to find
leading terms in the ALP-pion interactions, starting with the so-called
`mixing'-terms between the ALP and the neutral pions:
\begin{subequations}
    \label{eq:currmtch}
    \begin{align}
	    -i\frac{f_\pi^2}{4}\mathrm{Tr}
            \Big[\partial^\mu U_\pi^\dagger L_\mu U_\pi  \Big] + \text{h.c.}
            &= -\frac{1}{2}\frac{f_\pi}{f_a}C_L^8
            \partial^\mu a\, \partial_\mu \eta + \cdots \;,
        \\
	    i\frac{f_\pi^2}{4} \mathrm{Tr}
            \Big[\partial^\mu U_\pi^\dagger U_\pi  R_\mu  \Big] + \text{h.c.}
            &= \frac{1}{2}\frac{f_\pi}{f_a}C_R^3
            \partial^\mu a\:\partial_\mu \pi^0
            + \frac{1}{2}\frac{f_\pi}{f_a}C_R^8
            \partial^\mu a\:\partial_\mu\eta + \cdots \;,
        \\
	    \frac{\Lambda f_\pi^2}{2}\: \mathrm{Tr}
            \Big[\overline{M} U^\dagger_\pi\Big]+\text{h.c.}
        &= \frac{B_0 f_\pi}{f_a} \Bigg[
            \Bigg(
            \frac{C_{LR}^8}{\sqrt{3}}
            m_\Delta-C_{LR}^3\hat{m}
            \Bigg) a\pi^0
        \notag\\
            -\frac{1}{\sqrt{3}}\Bigg(&
            C_{LR}^3m_\Delta +
            \frac{C_{LR}^8}{\sqrt{3}}
            \hat{m}
        +
        \frac{2}{\sqrt{3}}C_{LR}^8
        m_s \Bigg)a\eta + \cdots \Bigg]\;,
       \\
       \mathrm{where}\; m_\Delta= \frac{m_u-m_d}{2},\quad
       &\hat{m}= \frac{m_u+m_d}{2},\quad
       \mathrm{and}\quad B_0=\Lambda.
    \end{align}
\end{subequations}
Note the absence of ALP mixing with $K^0$ and $\bar{K}^0$. Even though
electric charges allow for it, the ALP interactions in the quark mass
basis effectively preserve strangeness, since, $a$-$K^0$ or
$a$-$\bar{K}^0$ mixing terms are absent.  Further, 
both kinetic and mass mixing terms are suppressed by the factor
\(\xi\equiv f_\pi/f_a\),  which we use as the universal power counting
parameter relevant to the physics of the ALP. This suggests that one way
to calculate observables would be to treat the mixing terms as
interactions. However, in this work we take the more straightforward
approach of redefining the fields such that the Lagrangian is written in
terms of canonically normalized mass eigenstates.  The redefinition
first brings the kinetic terms in the canonically normalized form
followed by another rotation (orthogonal transformation) which
diagonalizes the squared mass matrix.  From this point, we would refer to
the basis in which \cref{eq:eq2.15,eq:eq2.16,eq:currmtch} are given as
the \textit{original-basis}, whereas the basis in which the kinetic
terms
are orthonormal and masses diagonal  would be referred to as the
\textit{eigenbasis}.

We begin with the kinetic mixing terms. The following field
redefinitions for \(a,\pi^0\), and \(\eta\) get rid of the kinetic
mixing terms at the order quadratic in \(\xi\):
\begin{align}
	\label{eq:eq2.22}
    \begin{split}
        a&\rightarrow a\;\left[
            1-\frac{\xi^2}{8}
            \left\{\left(C_R^3\right)^{2}
            +\left({C_A^8}\right)^{2}\right\}
        \right]
        - \pi^0\;\frac{\xi}{2} {C_{R}^3} - \eta\; \frac{\xi}{2} {C_{A}^8}
        \,;   \\
		\pi^0&\rightarrow
            \pi^0\!\left[
                1 + \frac{\xi^2}{8} \left({C_{R}^3}\right)^2
            \right]
            + \eta\; \frac{\xi^2}{4} {C_{R}^3} {C_{A}^8} \, ; \\
            \eta&\rightarrow \eta\,\, \left[
                1 + \frac{\xi^2}{8}\left({C_{A}^8}\right)^2
        \right],\quad \text{where} ~C_{A/V}^8 = C_R^8 \mp
        C_L^8.
    \end{split}
\end{align}
In addition to the kinetic mixing with the \(\pi^0\), \(\eta\), the
diagonal ALP kinetic term is also rescaled in the original basis. The
rescaling is sourced by the \( \mathrm{Tr}[L_\mu L^\mu+R_\mu R^\mu]\)
term given by the gauged kinetic term corresponding to \(U_\pi\)in
\cref{eq:eq2.16}:
\begin{align}
    \label{eq:eq2.16_ext}
    \mathrm{Tr}[L_\mu L^\mu+R_\mu R^\mu]\supset 
        \frac{1}{8}\left[(C_R^3)^2 + (C_A^8)^2 \right].
\end{align}
The redefinitions of the fields, as given in \cref{eq:eq2.22} reflect
this scaling of the ALP kinetic term in the original basis. Said 
redefinitions change the squared mass matrix for the \(a, \pi^0\), and
\(\eta\) fields to:
\begin{align}
    \begin{split}
        &\frac{1}{2}
        \begin{pmatrix} a &  \pi^0 &  \eta   \end{pmatrix}
    \begin{pmatrix}
        m_a^2 &  m_{a\pi}^2  &  m_{a\eta}^2 \\
	    m_{a\pi}^2 &  m^2_\pi &  m_{\pi\eta}^2 \\
	    m_{a \eta}^2 & m_{\pi\eta}^2 & m_{\eta}^2
    \end{pmatrix}
    \begin{pmatrix} a  \\  \pi^0 \\ \eta   \end{pmatrix}, \,
    \mathrm{where}\\
    m_{\pi}^2&= 2B_0\hat{m}\left[
        1 - \xi^2\frac{C_R^3}{4}\left(
        2 C_{LR}^3 - C^3_R +\frac{2}{\sqrt{3}} C_{LR}^8
        \frac{m_\Delta}{\hat{m}}\right)
    \right]+\cdots\;;
    \\
    m_{\eta}^2 &=
        \frac{2B_0}{3}(\hat{m}+2m_s) \left[
        1 + \frac{\xi^2}{4}C_{\!A}^8\left(
        {C_{\!A}^8}-2 C_{LR}^8
    \right)\right]+\cdots\;;
    \\
	m_{\pi\eta}^2&=
        \frac{2B_0}{\sqrt{3}}m_\Delta\left[
        1  - \frac{m_s}{m_\Delta} \xi^2
        \frac{C_R^3}{2}
        \frac{C_{LR}^8}{\sqrt{3}}
    \right] + \cdots\;;
    \\
    m_{a\pi}^2&= \xi B_0 \left[
        \frac{C_{LR}^8}{\sqrt{3}}
         m_\Delta + C_{LR}^3 \hat{m}
    \right] + \cdots\;;
    \\
    m_{a\eta}^2&= \xi \frac{B_0}{\sqrt{3}}\left[
        C_{LR}^3 m_\Delta
        + \frac{C_{LR}^8}{\sqrt{3}}
        (\hat{m}+2m_s)
     \right]+\cdots\;.
    \end{split}
\end{align}
Here \(\cdots\) represents higher order terms suppressed by additional
powers of \(\xi\),  \(m_{u/d}/m_s\), \(m_a^2/B_0m_s\)  etc. Before
proceeding, we need to comment about the sizes of these quantities and
the order of precision of our calculations.  In the SM\(\chi\)PT, the
only relevant mixing is between $\pi^0$ and $\eta$, which is rather
trivial to take into consideration since the mixing (parameterized by
$\epsilon$) is of the order of  $1 \times 10^{-2}$.  In this paper, we
only consider tree level effects linear in $\epsilon$, which implies
that we need to keep terms of the order of $\xi^2$ when scanning for
$\xi \lesssim \sqrt{\epsilon}$.  Additionally, we will also assume
$m_a^2 /B_0\hat{m} \lesssim  m_\pi^2 / m_\eta^2 \sim \epsilon$. This
suggests, we keep terms of the  order  of $\xi^2, m_a^2 /m_\pi^2$ but
neglect terms of the
order of $\xi \epsilon, m_a^2 /B_0 m_s$ etc. To keep the text, somewhat,
notationally simple, we represent \emph{all} contributions to the ALP mass,
i.e. those from the UV and those from QCD, by \(m_a\). 

An implication of these hierarchies is that the off-diagonal elements in the
mass-squared matrix are smaller than the differences of the
corresponding diagonal ones. Consequently, we can employ results from
non-degenerate perturbation theory to find the eigenvectors and
eigenvalues. Note that we are interested in determining the eigenvalues
and eigenvectors at order $\xi^2$. Apart from the perturbative
corrections, we also include the effects from redefinitions  in
\cref{eq:eq2.22}, which allows us to write the fields in the
original-basis  in terms of the fields  in the eigenbasis (denoted here
by hatted fields) as:
\begin{align}
    \label{eq:FlavinEigen}
    \begin{split}
        \pi^0&=
            \left(1
                - \frac{\xi^2}{8}\left(
                {C_{LR}^3}^2 - {C_R^3}^2
            \right)
        \right) \hat{\pi}^0
        +\left(\epsilon+\frac{\xi^2}{4}C_8C_{R}^3
    \right)\hat{\eta}
    - \xi\frac{C_{LR}^3}{2}
     \hat{a}\,;\\
    {\eta}&=
            \left(1
                - \frac{\xi^2}{8}\left(
                    {C_{LR}^8}^2-{C_{A}^8}^2
            \right)
        \right) \hat{\eta}
        -\left(\epsilon+\frac{\xi^2}{4}C_{LR}^8 C_3
    \right)\hat{\pi}^0
    -  \xi \frac{C_{LR}^8}{2}\:  \hat{a}\,; \\
		a&= \left(1
            - \frac{\xi^2}{8} \left(
                C_8^2 + C_3^2
            \right)
        \right)\hat{a}
        + \frac{\xi}{2} C_3 \hat{\pi}^0
        + \frac{\xi}{2} C_8  \hat{\eta}\,,
        \\
        &\text{where }
		\epsilon= \frac{\sqrt{3}}{4} \: \frac{m_u -m_d}{m_s -
        \hat{m}};\; C_{3} = C_{LR}^{3}-C_R^{3};\; 
        C_{8} = C_{LR}^{8}-C_A^{8}.
    \end{split}
\end{align}
The same procedure allows us to determine mass eigenvalues.  Below,
we summarize all the pion masses and include corrections due to
electromagnetism (denoted here by the quantity $\Delta_e$). For
completeness, we also include masses for charged  pions and neutral
kaons which remain the same as in the SM:
\begin{align}
	\label{eq2.25}
    \begin{split}
		M_{\pi^{\pm}}^2&= 2 B_0 \hat{m} + \Delta_e,\\
        M_{\pi^{0}}^2&= 2 B_0 \hat{m}\left[1 + \frac{1}{6}\left(
                3C_3^2 -2\sqrt{3} C_{LR}^8 C_R^3
                \frac{m_\Delta}{\hat{m}}
        \right) \xi^2
        \right]
        ,
        \\
        M_{K^{\pm}}^2&=  B_0 \left( m_s + m_u \right)+ \Delta_e,\\
        M_{K^{0}}^2&=  M_{\bar{K}^{0}}^2 = 	B_0
            \left( m_s + m_d\right),	\\
            M_{\eta}^2&=\frac{4}{3} B_0 \left(  m_s  + \frac{1}{2} \hat{m} \right)
        \left[1 + \frac{\xi^2}{4} C_8^2  \right].
    \end{split}
\end{align}

Hence, the spectrum of pions starts deviating from the SM
trajectories. This deviation of the trajectories of the meson masses allows us to constrain
\(\xi^2\) (multiplied by the relevant function of Wilson coefficients). In order to do so, we
construct the following sums out of the masses:
\begin{align}
       \Delta_{M_{\pi}}&\equiv \frac{M_{\pi^+}^2-M_{\pi^0}^2-\Delta_e}{M_{\pi^+}^2} = 
         0 + \frac{1}{6}\left(
           2\sqrt{3}C_{LR}^8C_R^3\frac{m_\Delta}{\hat{m}}-3C_3^2
         \right) \xi^2 +\cdots\;,\label{eq:mas_div1}\\
	\Delta_{\text{GMO}}&\equiv\frac{4 M_K^2 -M_\pi^2 - 3 M_\eta^2} {M_\eta^2 - M_\pi^2}
    =  0 - \frac{3}{4} \xi^2 C_8^2  + \cdots \; .\label{eq:mas_div2}
\end{align}
Here, we use the isospin invariant definitions $M_K^2 = \left(
M_{K^{0}}^2 +  M_{K^{\pm}}^2\right)/2$, and additionally use $M_\pi^2 =
M_{\pi^{\pm}}^2$. As always, the ellipsis represents terms further
suppressed.
The first sum is just a measure of the charged and neutral pion mass difference, while the second one
is the well known Gell-Mann--Okubo mass relation. 
Both the sums contructed are zero in the SM (up to \(\eta\) -- \(\eta^\prime\) mixing for GMO, which we discuss shortly). However, in the A\(\chi\)PT, the sum relations get violated at the tree level itself.

We can use the measured values of the meson masses and the computed values of the EM contributions in 
an attempt to give bounds on \(\xi^2\) from the mass differences. With \(M_{\pi^+}\) = 139.57039(18), 
\(M_{\pi^0}\)= 134.9768(5) \cite{Zyla:2020zbs}, and \(\sqrt{\Delta_e}\) = 4.538 MeV \cite{Donoghue:1996zn}, we have from \cref{eq:mas_div1}:
\begin{align}
    \label{eq:pidiff_num}
    2\sqrt{3}C_{LR}^8C_R^3\frac{m_\Delta}{\hat{m}}-3C_3^2 <0.38211(4).
\end{align}
In the isospin symmetric limit, \(m_\Delta=0\), the relationship is trivially satisfied. However, 
for \(m_\Delta\neq 0\), the relationship above constrains \(\xi^2\) times the Wilson coefficients.
However, in this case, it's contaminated by the presence of the quark masses, making the conclusion
inconclusive. 

The GMO relation, on the other hand, is constructed in a way that doesn't involve the quark masses. Hence, it can be used to bound \(C_8^2\xi^2\). In order to do so, however, we need to carefully take into account 
\(\eta\)--\(\eta^\prime\) mixing\footnote{We
        ignore the EM corrections to the charged meson masses and
    \(\eta^\prime\) mixing with the other neutral states as these are
    subdominant compared to \(M_\eta\) and \(\eta\)--\(\eta^\prime\) mixing,
    respectively. }. 
For the mass-mixing, we have:
   \begin{align}
       \label{eq:8}
       M_\eta^2&= \underbar{M}_\eta^2\cos^2\vartheta_{\eta\eta^\prime}
            + \underbar{M}_{\eta^\prime}^2\sin^2\vartheta_{\eta\eta^\prime}
   \end{align}
   Here, \(M_\eta\) is the mass of the \(\eta\) as given by
   \(SU(3)_L\times SU(3)_R\) chiral perturbation theory, while,
   \(\underbar{M}_\eta\) = 547.862(17) MeV and
   \(\underbar{M}_{\eta^\prime}\) = 957.78(6) \cite{Zyla:2020zbs} are the physical masses
   of the \(\eta\) and the \(\eta^\prime\) mesons, respectively. We use
   the lattice value \(\vartheta_{\eta\eta^\prime}=-14.1(2.8)^\circ\) \cite{Christ:2010dd, Bali:2021vsa}
   for the \(\eta\)--\(\eta^\prime\) mixing angle. With the magnitudes
   of the different quantities in hand, we get for the LHS of Eq
   \cref{eq:mas_div2}: \(\Delta_\mathrm{GMO}=-0.15\pm 0.13\). This translates to
   a bound on \(C_8^2\xi^2\),
   \begin{align}
       \label{eq:}
       C_8^2\xi^2<0.14
   \end{align}
   at 95\% CL. The, somewhat, weak bounds on the ALP parameters from the
   meson mass relationships have an interesting consequence. Quark
   masses are obtained on the lattice from the meson masses and the
   relationship between them (see, e.g., \cite{Aoki:2013ldr}), using the SM-only hypothesis. Hence, it
   is clear that these computed quark masses will deviate in the
   A\(\chi\)PT hypothesis. It is beyond the scope of this work to
   predict the exact nature of these changes.

For example, we see that the Gell-Mann--Okubo formula gets
violated at tree level (at $\mathcal{O}\left( \xi^2\right)$) even after
neglecting EM: 
\begin{equation}
	\Delta_{\text{GMO}}\equiv\frac{4 M_K^2 -M_\pi^2 - 3 M_\eta^2} {M_\eta^2 - M_\pi^2}
    \ = \  0 - \frac{3}{4} \xi^2 C_8^2  + \cdots \; ,
	\label{eq2.26}
\end{equation}
where we use the isospin invariant definitions $M_K^2 = \left(
M_{K^{0}}^2 +  M_{K^{\pm}}^2\right)/2$, and additionally use $M_\pi^2 =
M_{\pi^{\pm}}^2$. As always, the ellipsis represents terms further
suppressed.

Another important deviation from SM expectations happen for the meson
form factors (FF). Take, for example, the strangeness violating FFs
defined via
\begin{subequations}
    \label{eq:FFDef}
    \begin{align}
	    \langle \pi^0(p_\pi)|\bar{s}\gamma_\mu u|K^{+}(p_{K})\rangle
        &\equiv
        \frac{1}{\sqrt{2}} \Big[ f_{+,\,\mathrm{SM}}^{K^{+}
            \pi^{0}}(q^2)\; Q_\mu 
            + f_{-,\,\mathrm{SM}}^{K^{+}\pi^{0}}(q^2)\;q_\mu  
        \Big],
    \\
	    \langle \pi^{+}(p_\pi)|\bar{s}\gamma_\mu u|K^{0} (p_{K})\rangle
        &\equiv
	    f_{+,\,_\mathrm{SM}}^{K^{0} \pi^{-}}(q^2)\; Q_\mu
        +  f_{-,\,_\mathrm{SM}}^{K^{0} \pi^{-}}(q^2)\; q_\mu, \\
        \mathrm{where,}\quad Q^\mu= p_K^\mu+p_\pi^\mu;\quad& q_\mu=
        p_K^\mu-p_\pi^\mu.
    \end{align}
\end{subequations}
To find deviations in FFs in the A\(\chi\)PT, we match the
operator $\bar{s}_L\! \gamma_\mu u_L$ to pions in the original basis:
\begin{equation}
	\begin{split}
		\bar{s} \gamma_\mu u& = 
            - f_\pi^2 \text{Tr}\left[ 
            U_\pi^\dag  \left( t^4 - i t^5 \right) \partial_\mu U_\pi 
        \right]
        \\ & \supset
        \left( 
            K^{0}\partial_\mu \pi^{+} - \partial_\mu K^{0}\pi^{+} 
        \right) 
        + \frac{1}{\sqrt{2}} \left[ 
            K^{+} \partial_\mu  \left( 
                \pi_0 + \sqrt{3} \eta  
                \right) - \partial_\mu K^{+} \left( 
                \pi_0 + \sqrt{3}\eta  
        \right)  
    \right],
	\end{split}
\end{equation}
then, noting that the observed states get created/annihilated from/to
the vacuum by the fields (hatted) in the eigenbasis, we derive the form
factors (at the tree level from the $\mathcal{O}\left(p^2\right)$ 
Lagrangian) to find:
\begin{equation}
	\label{eq:FF_BSM}
	\begin{split}
        \frac{f_{+}^{K^{+} \pi^{0} }\left(0 \right)
        }{ f_{+}^{K^{0} \pi^{-}} \left(0 \right) }= 1 - \sqrt{3}
        \,\epsilon
        - \xi^2 \frac{C_3}{8}  \left[C_{A}^3+C_{LR}^3
        + 2\sqrt{3} C_{LR}^8 \right].
	\end{split}
\end{equation}
Here, \(f_{+}^{K\pi}(0)\) is the leading order term in the \(q^2\)
expansion of \(f_{+}^{K \pi}(q^2)\).  Compare the ratios of the form
factor as presented in \cref{eq:FF_BSM} to that of the SM
(\textit{i.e.}, in the limit $\xi^2 \rightarrow 0$).  The deviation from
the SM value of this ratio results due to the mixing of the ALP with the
\(\pi^0\). Similar deviations can be seen in the FFs corresponding to
other light mesons as well, \emph{e.g.},  \(f_{+}^{K^{+} \eta}(q^2)\),
\(f_{+}^{\pi^{+} \pi^0}(q^2)\) etc. As for the \(f_{-}^{K^{+}
\pi^{0}}(q^2)\) FFs, these are zero at leading order in the SM.
However, as we show in the next section, the ALP interactions 
source leading order contributions to these.  The modification of
the FFs, \(f_{\pm}^{K^{+} \pi^{0}}(q^2)\), are key elements in our discussion on
\(K^+\!\to\pi^0\ell^+\nu\) decays in the subsequent sections. 

\section{
    Phenomenology of \texorpdfstring{$K_{\ell_3}$}{Kl3} Decay 
    \label{sec:pheno}
}
The CC mediated semi-leptonic decays of the light mesons have always
been a testing ground for SM physics. The \(\pi^+\to\pi^0\ell^+\nu \)
(\(\pi_\beta\)) and the \(K^+\to \pi^0\ell^+\nu\) (\(K^+_{\ell_3}\))
decays are used to compute values of \(V_{\bar{d}u}\) and
\(V_{\bar{s}u}\) respectively. These decays are driven exclusively by
the \((\partial_\mu K^+(\pi^+) \pi^0 - K^+(\pi^+) \partial_\mu \pi^0)
\ell\gamma^\mu\bar P_L{\nu}\) operators at leading order in
SM\(\chi\)pt. However, in the A\(\chi\)PT, there are multiple terms that
are relevant for these decays, even at leading order. To proceed
further, we first list all the operators that generate contributions to
\(K^+_{\ell_3}\) at $\mathcal{O}(\xi^2)$.  Parametrizing the relevant
Lagrangian in the original-basis as:
\begin{equation}
	\mathcal{L}_{K_{\ell_3}^{+} } \ = \ 
    \sum_{i} \mathcal{C}_{K_{l_3}^{+}}^{i} 
    \mathcal{O}_{K_{l_3}^{+}}^{i},
	\label{eq:WilEx}
\end{equation}
we list the operators and the corresponding coefficients in
\Cref{tab:Ops} (with \(j_{-,\ell }^\mu=  \bar{\nu} \gamma^\mu
\frac{1}{2}(1-\gamma_5) \ell\)).    
\begin{table}[h]
	\centering
	\begin{tabular}{|l|l|}
    \hline
    ~&~\\
    \multicolumn{1}{|c|}{Operator \(\left(\mathcal{O}^i_{K^+_{\ell_3}}\right)\)}
    & \multicolumn{1}{c|}{Coefficient \(\left(\mathcal{C}^i_{K^+_{\ell_3}}\right)\)} \\
    ~&~\\
    \hline
    \hline
    &\\
    {$\mathcal{O}_{K_{\ell_3}^{+}}^0=   
        [ K^{+} \partial_\mu(\pi_0+\sqrt{3}\eta)
    $}
    & \multirow{2}{*}{$\mathcal{C}_{K_{\ell_3}^{+}}^0
    =i G_F V_{\bar{s}u}$} \\
    {$\qquad\qquad-\partial_\mu  K^+(\pi_0+\sqrt{3}\eta)] j_{-,\ell}^\mu$}
    & \\
    &\\
    \hline
    &\\
        $\mathcal{O}_{K_{\ell_3}^{+}}^1 =  
        \left( K^{+} \partial_\mu a - \partial_\mu K^{+} a \right)
        j_{-,\ell}^\mu$ 
        &
         $\mathcal{C}_{K_{\ell_3}^{+}}^1=
    i G_F V_{\bar{s}u}
    \dfrac{\xi}{2}
    (C_R^3 + \sqrt{3}C_R^8 - 2 i C_W )
    $
    \\
    &\\
    \hline
    &\\
    $\mathcal{O}_{K_{\ell_3}^{+}}^2 =
    \left(  K^{+} \partial_\mu a+ \partial_\mu K^{+}a\right) j_{-,\ell}^\mu$
    & $\mathcal{C}_{K_{\ell_3}^{+}}^2=i G_F V_{\bar{s}u}
    \dfrac{\xi}{2}
    ( C_R^3 + \sqrt{3} C_R^8 + 2 i C_W )
    $
    \\
    &\\
    \hline
    &\\
    $\mathcal{O}_{K_{\ell_3}^+}^3 = 
    \partial^\mu a \: \left( 
        \partial_\mu K^{+} K^{-} - K^{+} \partial_\mu  K^{-} \right)$ 
    & $\mathcal{C}_{K_{\ell_3}^{+}}^3=\dfrac{i}{4}\dfrac{1}{f_\pi}\xi\left(C_R^3+\sqrt{3}C_V^8\right)$\\
    &\\
    \hline
    &\\
        $\mathcal{O}_{K^+_{\ell_3}}^4 =
        \partial_\mu K^+  j^\mu_-$ 
    & $\mathcal{C}_{K_{\ell_3}^{+}}^4=-2 f_\pi\:G_F V_{\bar{s}u}$\\
    &\\
        \hline
	\end{tabular}
    \caption{ Operators in the \emph{original-basis} (before any mixing
        is taken into account) contributing to \(K^+_{\ell_3}\) decay at
        tree level. The operators in the eigenbasis are obtained by
        rotating the fields according to \cref{eq:FlavinEigen}.
    }
	\label{tab:Ops}
\end{table}

The operator $\mathcal{O}_{K_{l_3}^{+}}^0$ is the familiar operator of
the SM. However, as shown in the previous section, even this familiar
operator gives rise to deviations from the SM, due to the redefinition
of the physical pions.  The contributions due to  other operators  can
be calculated by converting fields in the original basis to the fields
in the eigenbasis by using the expansion given in \cref{eq:FlavinEigen}: 
\begin{subequations}
    \label{eq:OpFin}
\begin{align}
    \mathcal{C}_{K_{l_3}^{+}}^{1}\mathcal{O}_{K_{l_3}^{+}}^{1} 
     \!\supset
     \hat{\mathcal{C}}_{K_{l_3}^{+}}^{1}\hat{\mathcal{O}}_{K_{\ell}^{+}}^{1} 
    &\!=\! -i G_F V_{\bar{s}u} \xi^2 \frac{C_3}{4}
    [C_3+\!\sqrt{3}(C_8-C_L^8)\! + 2 i C_W]
        [\partial_\mu K^{+} \hat{\pi}^0\!   
            - K^{+} \partial_\mu  \hat{\pi}^0]\, 
      j_{-,\,\ell}^\mu,
        \\
     \mathcal{C}_{K_{l_3}^{+}}^{2}   \mathcal{O}_{K_{l_3}^{+}}^{2} 
     \!\supset
     \hat{\mathcal{C}}_{K_{l_3}^{+}}^{2}\hat{\mathcal{O}}_{K_{l_3}^{+}}^{2} 
    &\!=i G_F V_{\bar{s}u}\,
    \xi^2 \frac{C_3}{4}
        ( 
        {C_R^3+\sqrt{3}C_R^8}+ 2 i C_W)
         \partial_\mu (K^{+} \hat{\pi}^0 )
         j_{-,\,\ell}^\mu\;,
         \\
    \mathcal{C}_{K_{l_3}^{+}}^{3}\mathcal{O}_{K_{l_3}^{+}}^{3}\!\supset
    \hat{\mathcal{C}}_{K_{l_3}^{+}}^{3}\hat{\mathcal{O}}_{K_{l_3}^{+}}^{3}
    &\!=i\frac{\xi^2}{f_\pi}  
    \frac{C_3}{8}
        \left(C_R^3 + \sqrt{3}  C_V^8\right)
         \partial^\mu \hat{\pi}^0 \left(\partial_\mu  K^{+} K^{-} 
           -K^{+} \partial_\mu K^{-} \right)\; .
\end{align}
\end{subequations}
It is straightforward to see that \(\hat{\mathcal{O}}^1_{K^+_\ell}\)
gives a contribution proportional to \(Q^\mu\) to the matrix element,
while \(\hat{\mathcal{O}}^2_{K^+_{\ell_3}}\) gives a contribution
proportional to \(q^\mu\).  What is, perhaps, not as clear from the
operator itself is that the effective contribution of
\(\hat{\mathcal{O}}^3_{K^+_\ell}\) to the amplitude is proportional to
\(q^\mu\) as well.  This has been alluded to in \cref{sec:lag} and we
make it explicit here. The net contribution proportional to \(q_\mu\) is
given by: \(\mathcal{C}_{K_{l_3}^{+}}^4 \mathcal{C}_{K_{l_3}^{+}}^3 +
\mathcal{C}_{K_{l_3}^{+}}^2\). This implies that at tree level and at
\(\mathcal{O}(p^2,\epsilon,\xi^2)\),  the hadronic part of the amplitude
for \(K^\pm\to \pi^0\ell\nu\) gets additional contributions which are
not just limited to the modification of the FF $ f_{+}^{K^{+} \pi^{0}}
$.  Hence, we define the `effective' form factors \(\tilde{f}_\pm(0)\)
as:
\begin{align}
    \label{eq:ampDef}
        \mathcal{A} =  G_F  V_{\bar{s}u} &
		\left[ \tilde{f}_{+}^{K^{+} \pi^{0}}(0)\, Q_\mu +  
		\tilde{f}_{-}^{K^{+} \pi^{0}}(0)\, q_\mu  \right]  \: 
        \bar{u}_\nu \gamma^\mu \frac{1}{2}\left(1 - \gamma_5 \right) v_\ell.
\end{align}
After taking into account all operators listed in \Cref{tab:Ops}, we
find that:
\begin{align}
    \label{eq:albetDef}
    	&\tilde{f}_{+}^{K^{+} \pi^{0}}(0) = 
            \alpha^{(0)}_{K^+\pi_0}  + \xi^2
            \left(\alpha^{(2)}_{K^+\pi_0} 
                + i \tilde{\alpha}^{(2)}_{K^+\pi_0} \right)
            ;\quad 
        \tilde{f}_{-}^{K^{+} \pi^{0}} (0) = 
        \beta^{(0)}_{K^+\pi_0} +  \xi^2 
        \left(\beta^{(2)}_{K^+\pi_0} 
        + i \tilde{\beta}^{(2)}_{K^+\pi_0}\right),\notag \\
    &\text{with}~~
        \alpha^{(0)}_{K^+\pi_0}= 1 - \sqrt{3}\, \epsilon, 
        \;
        \alpha^{(2)}_{K^+\pi_0}= 
            -\frac{C_3}{8}( C_{LR}^3 - C_R^3 +2\sqrt{3} (C_{LR}^8 - C_R^8)),\;
        \tilde{\alpha}^{(2)}_{K^+\pi_0}= 
            -\frac{1}{2} C_3 C_W, \notag\\ 
    &\text{and}~~
        \beta^{(0)}_{K^+\pi_0}= 0,\;
        \beta^{(2)}_{K^+\pi_0}=-\frac{\sqrt{3}}{4}C_3 C_L^8,\;
        \tilde{\beta}^{(2)}_{K^+\pi_0}=\frac{1}{2}C_3 C_W.
\end{align}
Note, the imaginary parts of \(\tilde{f}_\pm^{K^{+} \pi^{0}} (0)\)
contribute to the decay amplitude only at \(\mathcal{O}(\xi^4)\). At
\(\mathcal{O}(\xi^2)\), only the \(\mathrm{Re}\left(
\tilde{f}_{\pm}^{K^{+} \pi^{0}}(0)\right)\) leave any signatures,
through interference with the leading SM contribution.  The definitions
of the FFs that we provide in \cref{eq:albetDef} make their extraction
from data a lot more straightforward.  The only catch is that the FFs
\(f_{\pm}^{K^{+} \pi^{0}}(0)\) defined in \cref{eq:FFDef} as matrix
elements of \(\bar{s}_L\gamma^\mu u_L\) is different from the effective
\(\mathrm{Re}\left( \tilde{f}_{\pm}^{K^{+} \pi^{0}}(0)\right)\) by
additional terms at \(\mathcal{O}(\xi^2)\).

There are additional contributions coming from higher orders in the
chiral expansion, from electromagnetism, and from EW breaking operators
\cite{Cirigliano:2001mk}.  One can simply absorb these by replacing $
\tilde{f}_\pm^{K^{+} \pi^{0}}(0) \rightarrow \tilde{f}_\pm^{K^{+}
\pi^{0}} (t)$ where $ t \equiv q^2$. Note, similar to the modifications
to the leading order form factors \(f_{\pm}^{K^+\pi^0}(0)\), as given in
\cref{eq:FF_BSM}, the strengths of these higher order contributions are
also expected to be modified.  Now, by the virtue of perturbativity of
the Lagrangian wrt \(\xi\), we can intuit: 
\begin{subequations}
    \label{eq:ffRedef}
\begin{align}
    \mathrm{Re}\left( \tilde{f}_{+}^{K^{+} \pi^{0}}(t)\right)& = 
        \left(
            \alpha^{(0)}_{K^+\pi^0}+\xi^2 \alpha^{(2)}_{K^+\pi^0}
            + \delta\alpha^{(0)}_{K^+\pi^0}
            + \xi^2 \delta \alpha^{(2)}_{K^+\pi^0}  
        \right) 
        \notag\\
        &\quad\times \Big[
            1 + \!\left(
            \lambda_{{K^+\pi^0}}^{+,\,(0)}  
            + \xi^2 \lambda_{{K^+\pi^0}}^{+,\,(2)} 
        \right)\!\frac{t}{M_\pi^2} 
        +\! \left(\lambda_{{K^+\pi^0}}^{\prime +,\,(0)} 
        + \xi^2 \lambda_{{K^+\pi^0}}^{\prime +,\,(2)} 
        \right)\!\frac{t^2}{2M_\pi^4}\Big]+ \cdots   \notag \\
        &\simeq
        \left[
            1 + \xi^2  
            \frac{\alpha^{(2)}_{K^+\pi^0}}{\alpha^{(0)}_{K^+\pi^0}}
        \!\right] f_{+,\,
            \mathrm{SM}}^{K^{+} \pi^{0}}(t)\;, \\
            \mathrm{Re}\left( \tilde{f}_{-}^{K^{+} \pi^{0}}(t)\right)& = 
        \left(
            \delta\beta^{(0)}_{{K^+\pi^0}}
            +\xi^2 \beta^{(2)}_{{K^+\pi^0}}
            +\xi^2 \delta \beta^{(2)}_{{K^+\pi^0}}  
        \right) \notag\\
        &\quad\times \Big[
            1 + \!\left(
            \lambda_{{K^+\pi^0}}^{-,\,(0)}  
            + \xi^2 \lambda_{{K^+\pi^0}}^{-,\,(2)} 
        \right)\!\frac{t}{M_\pi^2} 
        +\! \left(\lambda_{{K^+\pi^0}}^{\prime -,\,(0)} 
        + \xi^2 \lambda_{{K^+\pi^0}}^{\prime -,\,(2)} 
        \right)\!\frac{t^2}{2M_\pi^4}\Big]+ \cdots \notag\\
        &\simeq 
        \left[
            1 + \xi^2  
            \frac{\beta^{(2)}_{K^+\pi^0}}{\delta\beta^{(0)}_{K^+\pi^0}}
        \!\right] f_{-,\,
            \mathrm{SM}}^{K^{+} \pi^{0}}(t)\;,
\end{align} 
\end{subequations}
where \(\delta \alpha^{(0)}_{K^+\pi^0}\) and \(\delta
\alpha^{(2)}_{K^+\pi^0}\)  represent higher-order corrections
\cite{Cirigliano:2001mk}.  Out of these terms, we can neglect \(\xi^2
\delta \alpha^{(2)}_{K^+\pi^0}\) since we expect $\delta
\alpha^{(2)}_{K^+\pi^0} \lesssim \delta \alpha^{(0)}_{K^+\pi^0} \sim
10^{-2}$.  The \(\lambda^\pm\) are `slope parameters' that parametrize
the effects of the higher order  (in powers of $ q^2 $) terms in the
chiral expansion.  Obviously, these slope parameters get \(\xi^2\)
dependent contributions as well. Similar
arguments can be made for the parameters entering
\(\mathrm{Re}\;(\tilde{f}^{K^+\pi^0}_-(0))\). The only difference is
that in the SM there is no leading order contribution to
\(\mathrm{Re}\;(\tilde{f}^{K^+\pi^0}_-(0))\), making
\(\beta^{(0)}_{K^+\pi^0}=0\).  Putting everything together, the
spin-summed matrix element squared for \(K^+\to \pi^0\ell\nu\), at
$\mathcal{O}(\xi^2)$,  is given by:
\begin{align}
    \label{eq:finamp}
    \overline{|\mathcal{A}|}^2_{K_{l3}}&=
    2G_F^2 |V_{\bar{s}u}|^2 C_\mathrm{cor}  
        \left[
            1 + 2\, \xi^2  
            \frac{\alpha^{(2)}_{K^+\pi^0}}{\alpha^{(0)}_{K^+\pi^0}}
        \!\right]  
    (2H\cdot p_\ell\; H\cdot
    p_{\nu_\ell} - H^2p_\ell \cdot p_{\nu_\ell}),
    \\
     \mathrm{where}\; H_\mu&\equiv 
                  f_{+,\text{SM}}^{K^{+} \pi^{0}}(t)\, Q_\mu 
            +
       \left[
             1 + \xi^2\! 
             \left(\frac{\beta^{(2)}_{K^+\pi^0}}{\delta\beta^{(0)}_{K^+\pi^0}}
           -\frac{\alpha^{(2)}_{K^+\pi^0}}{\alpha^{(0)}_{K^+\pi^0}}
    \right)\!
        \right]f_{-,\text{SM}}^{K^{+} \pi^{0}}(t)\,q_\mu\notag.
\end{align}
The factor \(C_\mathrm{cor}\) encapsulates effects that are not captured
by the lattice computations of the FFs that we use in our numerical
analyses. It is defined as:
\begin{subequations}
\begin{align}
     C_\mathrm{cor}&= 
        S_{\rm EW}(1+\delta^K_{SU(2)}+\delta^{k\ell}_{\rm em})^2.
    \\\mathrm{Here,}\quad 
    S_{\rm EW}&= 1+ \frac{2\alpha_\mathrm{EM}}{\pi}\left(
    1-\frac{\alpha_s}{4\pi}\right)\log\frac{M_Z}{M_\rho}
    +\mathcal{O}\left(\frac{\alpha\alpha_s}{\pi^2}\right)
    = 0.0232\pm 0.0003,
\end{align}
encodes the short-distance contribution to
the EM corrections \cite{Sirlin:1981ie, FlaviaNet:2008hpm}.
The other corrections, \emph{viz.} \(\delta^{K\ell}_{SU(2)} =(2.45\pm0.19)\%\)
\cite{Moulson:2017ive} and \(\delta^{K\ell}_{\rm EM} =(0.016\pm0.25)
\times 10^{-2}\) \cite{FlaviaNet:2008hpm} are the isospin breaking and
long distance electromagnetic corrections respectively. In our
calculations, We drop the \(\delta^{K\ell}_{\rm EM}\) corrections as it
is much smaller than the order we are working up to.
\end{subequations}

\begin{figure}[htpb]
	\centering
    \includegraphics[scale=1]{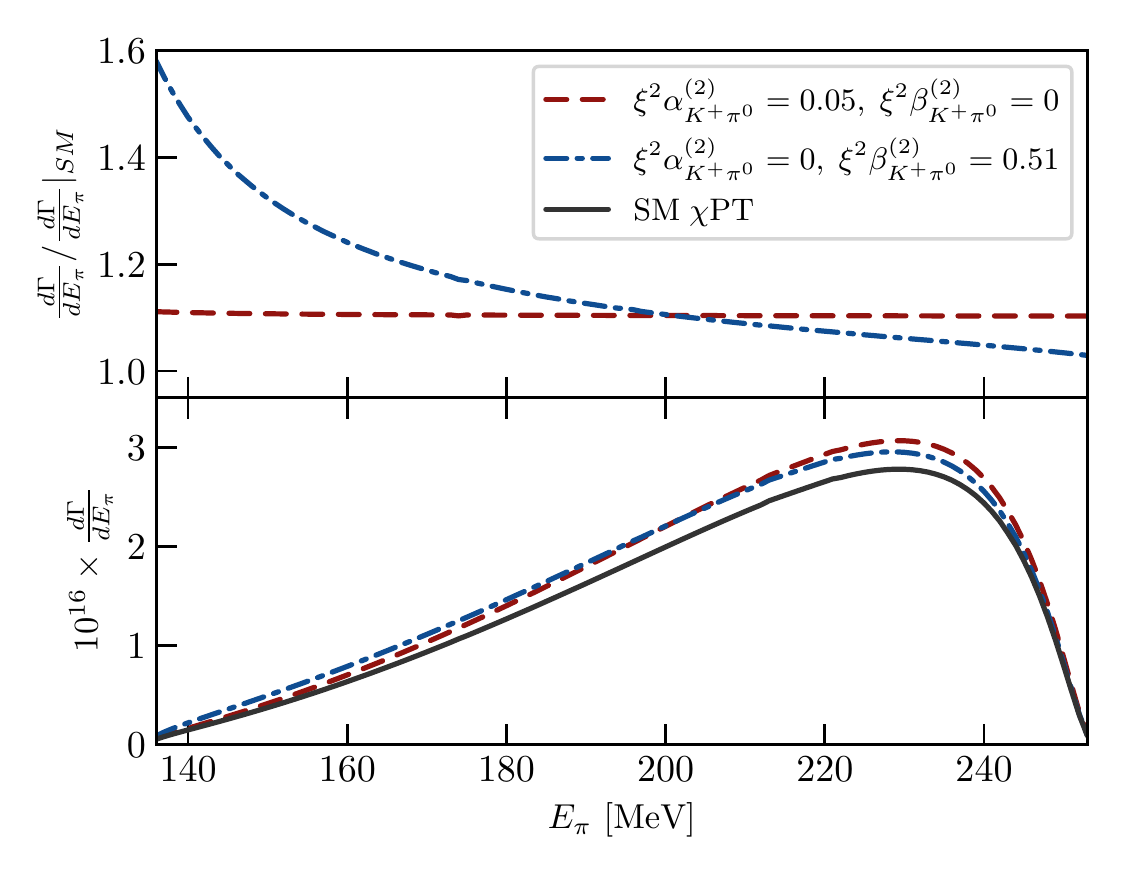}
    \caption{
        Bottom: Marginal spectrum of \(K^+_{\mu_3}\), wrt
        \(E_{\pi^0}\) for \(\xi^2\alpha^{(2)}_{K^+\pi^0}= 0.05\) (dashed
        red) and \(\xi^2 \beta^{(2)}_{K^+\pi^0} = 0.51\) (dash-dotted
        blue), are given.  The values of ALP parameters are chosen such
        that the total decay width is same for both graphs. For
        reference, the SM decay spectrum is shown in black. As is clear,
        to achieve same total decay rate, the required value of \(\xi^2
        \beta^{(2)}_{K^+\pi^0}\) is much larger than that of $ \xi^2
        \alpha^{(2)}_{K^+\pi^0} $ due to the lepton $ (\mu) $ mass
        suppression. Top: The same two NP spectra normalized by
        the SM spectrum.  This plot shows the different momentum
        dependences of the effects corresponding to the effective ALP
        parameters, $\xi^2 \alpha^{(2)}_{K^+\pi^0} $ and $\xi^2
        \beta^{(2)}_{K^+\pi^0} $. The values of the parameters are again
        chosen to keep the total width equal in both the cases.
    }
	\label{fig:alpha-beta-comp-wSM}
\end{figure}

As evident from \cref{eq:finamp}, there are two distinct NP effects to the 
amplitude-squared. The first is an overall scaling that is entirely controlled by
\(\alpha^{(2)}_{K^+\pi^0}\), and the second is a relative scaling between the two 
independent momentum directions.  
The relative scaling is proportional to the momentum 
transferred to the lepton system (\(q_\mu\)), and---with the leptons in the final state---it is essentially proportional to the mass of the charged lepton (\(m_\nu\simeq 0\)).
The net effect due to the relative scaling factor, being lepton-mass--suppressed, is then 
subdominant compared to that of the overall scaling term. 
Although, the relative scaling gets contributions from both \(\alpha^{(2)}_{K^+\pi^0}\) 
and \(\beta^{(2)}_{K^+\pi^0}\), the former is scaled by the leading order
\(\alpha^{(0)}_{K^+\pi^0}\), as compared to the
sub-leading \(\delta\beta^{(0)}_{K^+\pi^0}\) scaling of
\(\beta^{(2)}_{K^+\pi^0}\), implying that \(\beta^{(2)}_{K^+\pi^0}\) is the primary driver 
of the distortion in the decay distribution. Therefore, the primary effect of 
\(\alpha^{(2)}_{K^+\pi^0}\) is an overall scaling of the decay rate,
while those of \(\beta^{(2)}_{K^+\pi^0}\) are lepton-mass--suppressed and 
related to the distortion of 
the shape of the decay distribution.  In
\Cref{fig:alpha-beta-comp-wSM}, we show the effects of varying
\(\alpha^{(2)}_{K^+\pi^0}\) and \(\beta^{(2)}_{K^+\pi^0}\) on the
\(K^+_{\mu_3}\) marginal distribution wrt \(E_\pi\), obtained by
marginalising over \(E_\ell\) in:
\begin{align}
\label{eq:diffdecay}
\begin{split}
    \frac{d^2\Gamma}{dE_{\pi}dE_\ell}&= 
        \frac{1}{2\pi^3} \frac{1}{8{M_{K_+}}}
        \Theta(1-y^2)\Theta(M_{K_+}-E_{\pi}-E_\ell)
        \overline{|\mathcal{A}|}^2_{K_{\ell3}},\\
    \mathrm{with,}\;y&= 
        M_{K_+}^2 + M_{\pi}^2 + M_\ell^2 -2 M_{K_+} (E_{\pi}+E_\ell)
        + 2 E_{\pi}E_\ell,
\end{split}
\end{align}
where \(\Theta(x)\) is the Heaviside step function. From  the bottom
panel of \Cref{fig:alpha-beta-comp-wSM}, where we show the differential
distributions of $ K^+_{\mu_3} $, we can clearly see that compared to
\(\alpha^{(2)}_{K^+\pi^0}\), a large value of \(\beta^{(2)}_{K^+\pi^0}\)
is required to produce a similar change in the decay rate.  From the top
panel of the same figure, we can see the different momentum-dependence
of the modifications sourced by \(\alpha^{(2)}_{K^+\pi^0}\) and
\(\beta^{(2)}_{K^+\pi^0}\). 

\subsection{
    Results and constraints 
    \label{sec:RanC}
}
In the previous subsection, we derived the amplitude-squared for the
decay of \(K^+\to \pi^0\ell^+\nu\) in the A\(\chi\)PT. In this
subsection, we see how to constrain the effective parameters controlling
the deviations from the SM\(\chi\)PT, viz. \(\xi^2\alpha^{(2)}_{K^+\pi^0}\)
and \(\xi^2\beta^{(2)}_{K^+\pi^0}\), using data corresponding to
\(K^+_{\ell_3}\) decays. To constrain these parameters, we use the
following independent measurements:
\begin{itemize}
    \item Measurement of the differential decay distributions of
        \(K^\pm\to\pi^0\mu^\pm \nu_\mu (K^+_{\mu_3})\) and
        \(K^\pm\to\pi^0e^\pm \nu_e (K^+_{e_3})\) by the NA48/2
        collaboration at the CERN SPS \cite{Lazzeroni:2018glh}.
    \item The total width measurements for \(K^+_{\mu_3}\) and \(
        K^+_{e_3}\) decays. We have used the experimental averages of
        the branching fractions from PDG \cite{Zyla:2020zbs} to
        calculate the rates.
\end{itemize}
As discussed in the last subsection, the effect of \(\xi^2
\alpha^{(2)}_{K^+\pi^0}\) dominates over that of the
lepton-mass--suppressed effect of \(\xi^2 \beta^{(2)}_{K^+\pi^0}\) when
it comes to the total decay rate.  This suppression causes \(\xi^2
\beta^{(2)}_{K^+\pi^0}\) to be essentially unbounded by observables
related to \(K_{e_3}^+\).  However, as we show below, the marginal
energy spectra of the decay rate of \(K^+_{\mu_3}\) can be used
effectively to constrain \(\xi^2 \beta^{(2)}_{K^+\pi^0}\). 

As is evident from the discussion around \cref{eq:finamp}, the NP
effects will show up as deviations from the SM expectations of the FF
parameters. This implies that we need to carefully estimate the SM
inputs that enter \cref{eq:ffRedef}.  For the slope parameters,
\(\lambda_\pm\), appearing in \cref{eq:ffRedef}, we use results obtained
from lattice computations by the European twisted mass collaboration
\cite{Carrasco:2016kpy}. The collaboration reports the FFs,
expressed as:
\begin{align} 
    \label{eq:FFdisp} 
    \begin{split}
    f^{K^+\pi^0}_{+/0,\,\mathrm{SM}}(t)
    &= f^{K^+\pi^0}_{+/0,\,\mathrm{SM}}(0)
        \left[ 
       1 + \lambda_{K^+\pi^0}^{+/0,\,(0)}\frac{t}{M_\pi^2}
       + \frac{1}{2} \lambda_{K^+\pi^0}^{\prime\,+/0,\,(0)}
            \frac{t^2}{M_\pi^4}
    \right] +\cdots\;,\\
    f^{K^+\pi^0}_{-,\,\mathrm{SM}}(t)
    &= \left[
        f^{K^+\pi^0}_{0,\,\mathrm{SM}}(t)
        - f^{K^+\pi^0}_{+,\,\mathrm{SM}}(t)
    \right]{\dfrac{M_K^2 - M_\pi^2}{t}}\;.
\end{split}
\end{align}
We use the dispersive parametrization of the FFs, where the slope
parameters are determined in terms of the slope of the vector,
\(f^{K^+\pi^0}_{+}(t)\), FF at \(t=0\) and the slope of the
scalar, \(f^{K^+\pi^0}_{0}(t)\), FF at the Callan-Treiman point
(\(t_{CT}= M_K^2-M_\pi^2\)), \(\Lambda_+\) and \(C\) respectively
\cite{Bernard:2009zm}: 
\begin{align}
    \label{eq:FF_param}
    \begin{split}
    \lambda_{K^+\pi^0}^{+\,(0)}&= \Lambda_+\;;\\
    \lambda_{K^+\pi^0}^{'+\,(0)}& = 
        \left(\lambda_{K^+\pi^0}^{+\,(0)}\right)^2 + 5.79(97)\times 10^{-4}\;;\\
    \lambda_{K^+\pi^0}^{0\,(0)}&= 
        \dfrac{M_\pi^2}{t_{CT}}\left[\log(C)- 0.0398(44)\right]\;;\\
    \lambda_{K^+\pi^0}^{'0\,(0)}& = 
        \left(\lambda_{K^+\pi^0}^{0\,(0)} \right)^2 + 4.16(56)\times 10^{-4}.
    \end{split}
\end{align}
We tabulate the lattice determination~\cite{Carrasco:2016kpy} of $
\Lambda_+, C $, and $f^{K^+\pi^0}_{+,\,\mathrm{SM}}(0) (=
f^{K^+\pi^0}_{0,\,\mathrm{SM}}(0))$ in \Cref{tab:SMFFparameters}. We
also note the correlations among these parameters, which we include in
our computations.

\begin{table}[htpb]
	\centering
	\renewcommand{\arraystretch}{1.3}
	\begin{tabular}{|l|l|}
		\hline
		Parameter & Correlation \\
		\hline
		\hline
		$\Lambda_+ = 24.22 (1.16) \times 10^{-3}$ 
            & $\rho\left[\Lambda_+, \log(C)\right] = 0.376$\\
		\hline
		$\log(C) = 0.1998 (138)$ 
            & $\rho\left[f^{K^+\pi^0}_{+/0,\,\mathrm{SM}}(0), \log(C)\right] 
                = -0.719$\\
		\hline
		$f^{K^+\pi^0}_{+/0,\,\mathrm{SM}}(0) = 0.9709 (46)$ & $\rho\left[f^{K^+\pi^0}_{+/0,\,\mathrm{SM}}(0), \Lambda_+\right] = -0.228
		$\\ 
		\hline
	\end{tabular} 
    \caption{Lattice determined values (left) and correlations (right)
        of the FF parameters in the dispersive formalism
        \cite{Carrasco:2016kpy}. The quantities $ \rho[i,j] $ are the
        correlation coefficients between the $ i $-th and the $ j $-th
        parameters. 
    }
	\label{tab:SMFFparameters}
	\renewcommand{\arraystretch}{1}
\end{table}

We have checked that the contributions from
\(\lambda^{\prime\prime +/0,\,{(0)}}_{K^+\pi^0}t^3/M_\pi^6\) and higher
are at a sub percent level, much smaller than the experimental
precision. The smallness of \(\lambda^{\prime\prime\, +/0}_{K^+\pi^0}\)
is why we have truncated \(f_{-,\,\mathrm{SM}}^{K^+\pi^0}(t)\) after
the \(t/M_\pi^2\) term, even though  \(t/M_\pi^2\) can be as large as
\(\sim\) 10 near the edge of the phase space.  As is obvious from
\cref{eq:FFdisp}, the
coefficient of the \(t^2\) term would be proportional to
\((\lambda^{\prime\prime +\,,{(0)}}_{K^+\pi^0}-\lambda^{\prime\prime\,
0\,,{(0)}}_{K^+\pi^0})\), which is below the experimental sensitivity.  

Another subtlety to be considered in the analyses is the value of
\(V_{\bar{s}u}\). The $K^+_{\ell_3}$ total width measurements are often
used to independently determine the CKM element $V_{\bar{s}u}$
\cite{Zyla:2020zbs}.  We can't use \(V_{\bar{s}u}\), extracted from
\(K^+_{\ell_3}\) under the SM only hypothesis, as an independent
parameter  while fitting the NP hypothesis to the same \(K^+_{\ell_3}\)
data.  To circumvent this issue, we need to use a determination of
\(V_{\bar{s}u}\) that does not use the \(K^+_{\ell_3}\) data at all.
Such a computation does exist, where the ratio of the \(K^+\to
\mu^+\nu\) width to the \(\pi^+\to \mu^+\nu\) width is used to obtain
$V_{\bar{s}u} = 0.2252 \pm 0.0005$ ~\cite{Marciano:2004uf,
Zyla:2020zbs}. This extraction is suitable for the analysis as none
of these amplitudes are modified in the A\(\chi\)PT at
\(\mathcal{O}(\xi^2)\).

\begin{figure}[htpb]
    \centering
	\includegraphics{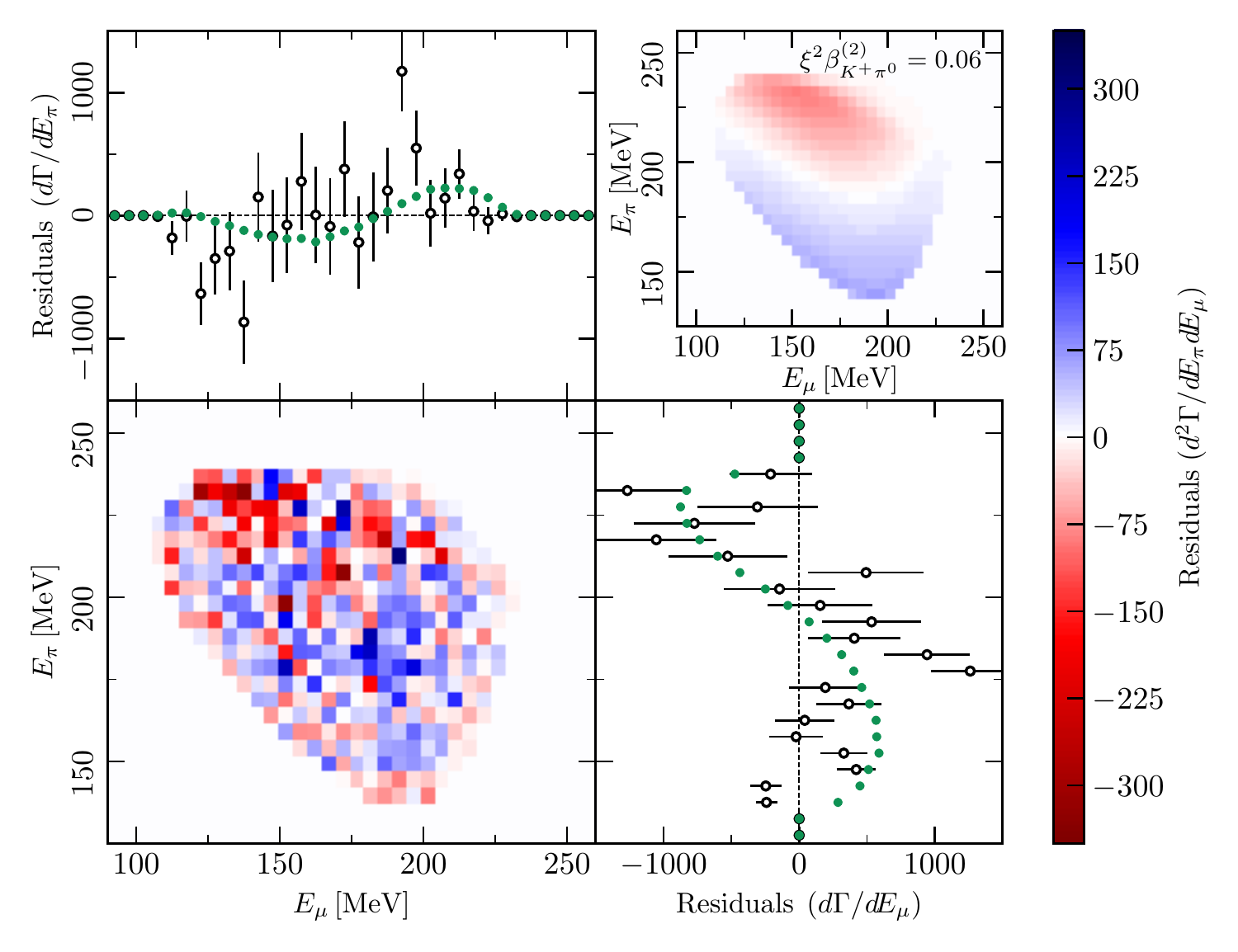}
    \caption{Bottom Left: The binned Dalitz plot corresponding to the
        residuals, i.e., the difference between the collected data
        and the theoretical SM predictions.  Sharing its x- and y-axes
        to the top and to the right are the corresponding marginals with
        respect to \(E_\mu\) and \(E_\pi\). The empty circles denote the
        residual signal and the bars denote the 1\(\sigma\) experimental
        error. Top Right: The binned Dalitz plot corresponding to the
        additional events generated by NP for \(\xi^2
        \beta^{(2)}_{K^+\pi^0} = 0.06\). The corresponding marginals are
        depicted in the panels containing the marginals of the residuals
         for comparison (Green filled circles). For effective comparison, we
        use the same normalisation of color-shading for the data and the
        theory distributions.}
	\label{fig:exp}
\end{figure}

On the experiment side, we use the \(K^+_{\ell_3}\) data obtained by the
NA48/2 collaboration\footnote{Publicly available at:
\url{https://zenodo.org/record/3560600\#.X-xCjulKjUJ}
\cite{zenodona4818} (CC BY 4.0).} to fit the truth level distribution
against the observed distribution. We combine this multi-variable fit
with the constraints set by the independent measurements of the total
decay rates to bound \(\xi^2 \alpha^{(2)}_{K^+\pi^0}\) and \(\xi^2
\beta^{(2)}_{K^+\pi^0}\).  The data consist of bin-by-bin event
distributions of the differential decay rate with respect to the pion
and the lepton energies \((E_\pi, E_\mu)\), for \(4.4 \times 10^6\) and
\(2.3 \times 10^6\) reconstructed events corresponding to \(K^+_{e_3}\)
and \(K^+_{\mu_3}\) respectively.  Using the data for \(K^+_{\mu_3}\),
we draw the binned Dalitz distribution for the residual events, defined
as the differences between accepted events and SM
predictions. We show this in the bottom-left panel of \Cref{fig:exp}.
The diagonal panels show the distributions of excess events with
\(1\sigma\) experimental error-bars with respect to \(E_{\pi}\)
(bottom-right) and \(E_\mu\) (top-left), after marginalizing over the
\(E_\mu\) and \(E_{\pi}\) bins respectively.  On the top-right panel, we
show the 2D distribution of excess NP events with \(\xi^2
\beta^{(2)}_{K^+\pi^0} = 0.06,\; \xi^2\alpha^{(2)}_{K^+\pi^0}=0\).  The
binned and marginalized distributions of the excess BSM events are shown
in the panels containing the corresponding marginal distributions for
the data. Although the residual fluctuations show a slight systematic
excess in \Cref{fig:exp}, the excess becomes consistent with the SM
prediction when the theory error
is taken into account.

\begin{figure}[htpb]
	\centering
	\includegraphics{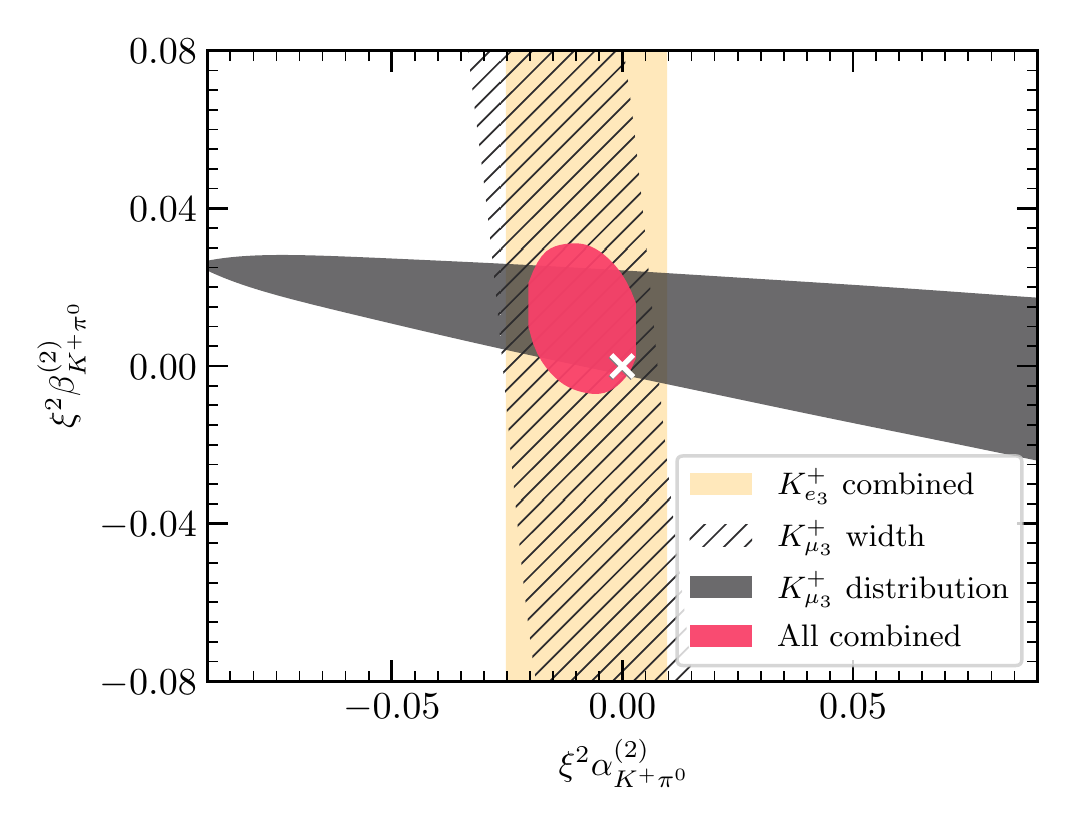}
    \caption{The \(95\%\) C.L. allowed regions for the ALP parameters in
        the \(\xi^2 \alpha^{(2)}_{K^+\pi^0}\) -
        \(\xi^2\beta^{(2)}_{K^+\pi^0}\) plane. The yellow band indicates
        the region allowed by the combined \(K^+_{e_3}\) data,
        \emph{i.e.}, the total rate and differential rates combined.  The
        black patch shows the region allowed (at 95\% C.L.) by the
        \(K^+_{\mu_3}\) differential distribution. The hatched area is
        the corresponding region allowed by the \(K^+_{\mu_3}\) total
        rate. The red patch is the 95\% C.L. allowed region obtained by
        combining all the independent analyses. The cross marks the
        SM\(\chi\)PT point where the values of both the parameters are
        zero.
    }
	\label{fig:alpha-beta-contour}
\end{figure}

The data corresponding to the differential rate for \(K^+_{\mu_3}\) is
the primary source of constraint for \(\xi^2\beta^{(2)}_{K^+\pi^0}\).
We note, both \(\xi^2\beta^{(2)}_{K^+\pi^0}\) and
\(\xi^2\alpha^{(2)}_{K^+\pi^0}\) appear in the same footing as the
relative factor between \(q^\mu\) and \(Q^\mu\) in the amplitude given
in \cref{eq:finamp}. The exact form of this factor is: 
\begin{equation*}
    \left(\dfrac{\xi^2\beta^{(2)}_{K^+\pi^0}}{\delta\beta^{(0)}_{K^+\pi^0}}
    -\dfrac{\xi^2\alpha^{(2)}_{K^+\pi^0}}{\alpha^{(0)}_{K^+\pi^0}}
    \right).
\end{equation*}
However, the bound on \(\xi^2\beta^{(2)}_{K^+\pi^0}\) from the
differential distribution is stronger than that on
\(\xi^2\alpha^{(2)}_{K^+\pi^0}\). This is because
\(\xi^2\beta^{(2)}_{K^+\pi^0}\) and \(\xi^2\alpha^{(2)}_{K^+\pi^0}\) are
scaled by factors that are hierarchically separated, with
\(|\delta\beta^{(0)}_{K^+\pi^0}|\sim 0.1 \times
|\alpha^{(0)}_{K^+\pi^0}|\). Thus, decay rate and decay distribution
measurements provide complementary constraints on these two parameters.

Before proceeding to a more sophisticated analysis to constrain
\(\xi^2\beta^{(2)}_{K^+\pi^0}\), it is instructive to get a heuristic
estimate for the order of the constraint we can achieve. Note that, the
NA48/2 collaboration fits the measured differential distribution
assuming the SM\(\chi\)PT and determines the values of the slope
parameters \cite{Lazzeroni:2018glh}.  We can get an estimate of
\(\xi^2\beta^{(2)}_{K^+\pi^0}\) from the difference of the measured
(fitted) and the SM computations of the slope parameters. Using
\cref{eq:FFdisp} and \cref{eq:ffRedef}, we can find an approximate
expression for \(\xi^2 \beta^{(2)}_{K^+\pi^0}\):
\begin{align}
    \label{eq:rough}
    \xi^2 \beta^{(2)}_{K^+\pi^0}&\approx 
    f_{-,\,\mathrm{Fit}}^{K^+\pi^0}(0) - \delta\beta^{(0)}_{K^+\pi^0}
        \equiv f_{-,\,\mathrm{Fit}}^{K^+\pi^0}(0) 
        - f_{-,\,\mathrm{SM}}^{K^+\pi^0}(0) \notag\\
    &\approx \frac{M_K^2 - M_\pi^2}{M_\pi^2}\left[\!
       \left (\lambda_{K^+\pi^0}^{+,\,(0),\mathrm{Fit}}
       - \lambda_{K^+\pi^0}^{0,\,(0),\mathrm{Fit}}\right)
      \! - f_{+,\,\mathrm{SM}}^{K^+\pi^0}\!(0)\left(
       \lambda_{K^+\pi^0}^{+,\,(0),\mathrm{SM}}
       - \lambda_{K^+\pi^0}^{0,\,(0),\mathrm{SM}}\right)
       \!\right]\!,
\end{align}
if we take \(\xi^2\alpha^{(2)}_{K^+\pi^0}\to 0\).  Using the SM values
from \Cref{tab:SMFFparameters} and fitted numbers from Table 4 of Ref.
\cite{Lazzeroni:2018glh}, \cref{eq:rough} roughly yields
\(\xi^2\beta^{(2)}_{K^+\pi^0}\sim 0.01 \pm 0.04\).  As we show now, a
more involved computation using simultaneous fits yields a constraint of
a similar size. 

We compute \(\chi^2\) distributions by comparing the truth-level signal
against the differential distribution data and the total decay width
measurement, after taking into account the experimental and theoretical
errors and correlations. For the differential distributions, we
normalize our histograms using the total number of events, as quoted in
the last paragraph.  In \Cref{fig:alpha-beta-contour}, we show the
\(95\%\) confidence limits (C.L.) obtained in our analysis in the
(\(\xi^2 \alpha^{(2)}_{K^+\pi^0}\) - \(\xi^2\beta^{(2)}_{K^+\pi^0}\))
plane.  The yellow band shows the combined exclusion obtained from the
differential and total decay rate measurements of \(K^+_{e_3}\). Due to
the smallness of the electron mass, the constraint on
\(\xi^2\beta^{(2)}_{K^+\pi^0}\) is insensitive to the decay distribution
measurement.  For \(K^+_{\mu_3}\), we show the individual exclusions
obtained from the differential and the total decay rate data, the solid
black and the hatched regions respectively. In analysing the figure, we
clearly see that the \(K^+_{\mu_3}\) total rate measurement (hatched)
mostly constrains \(\xi^2 \alpha^{(2)}_{K^+\pi^0}\). Whereas, the
differential distribution (black) constraints primarily \(\xi^2
\beta^{(2)}_{K^+\pi^0}\), as anticipated.  The black patch is not
centred around the SM point
\((\xi^2\alpha^{(2)}_{K^+\pi^0}=\xi^2\beta^{(2)}_{K^+\pi^0}=0)\), but is
compatible with it at \(2\sigma\).  This is because there is a slight
disagreement between the SM and the measured values of the FF parameters
(see Figure 8 of Ref.~\cite{Carrasco:2016kpy}).  The combined exclusion,
from the four independent measurements is shown in red. In
\Cref{tab:beta-alpha-fixed}, we show the individual constraints by
varying one parameter at a time. We can see from the table that the
addition of the \(K_{e_3}^+\) measurements only improves the constraint
on \(\alpha^{(2)}_{K^+\pi^0}\).

\begin{table}[htpb]
	\centering
    \renewcommand{\arraystretch}{1.4}
	\begin{tabular}{| c| c | c| }
		\hline
        &  \(\xi^2\beta^{(2)}_{K^+\pi^0}\; \left(\xi^2 \alpha^{(2)}_{K^+\pi^0} =
        0\right)\) & \(\xi^2\alpha^{(2)}_{K^+\pi^0}\;
        \left(\xi^2\beta^{(2)}_{K^+\pi^0} = 0\right)\)\\
		\hline
		\hline
		\(K_{\mu_3}^+\) & $ -0.006<\xi^2\beta^{(2)}_{K^+\pi^0}< 0.026$  &    $ -0.021<\xi^2\alpha^{(2)}_{K^+\pi^0}< 0.007$\\
		\hline
		\(K_{\mu_3}^++K_{e_3}^+\) & $ -0.006<\xi^2\beta^{(2)}_{K^+\pi^0}<0.026 $& $    -0.018<\xi^2\alpha^{(2)}_{K^+\pi^0}<0.003$\\
		\hline
	\end{tabular}
    \caption{Tabulated here are the  95\% confidence limits on 
        \(\xi^2\alpha^{(2)}_{K^+\pi^0}\) and 
        \(\xi^2\beta^{(2)}_{K^+\pi^0}\) from the \(K_{\mu_3}^+\) decay analysis and the $K_{\mu_3}^++K_{e_3}^+$
        combined analysis.
        }
	\label{tab:beta-alpha-fixed}
    \renewcommand{\arraystretch}{1}
\end{table}

\begin{figure}[ht]
	\centering
	\includegraphics{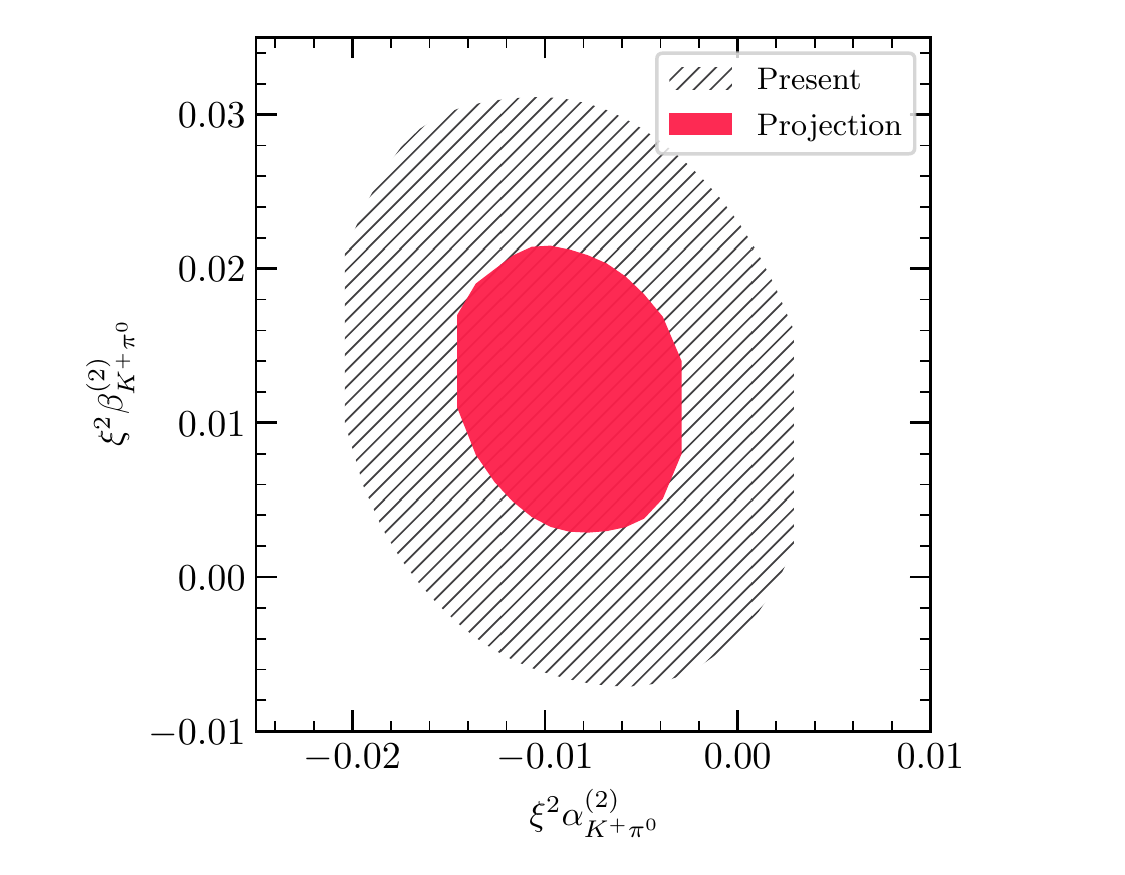}
    \caption{A conservative future projection (red) for 95\% C.L.
        exclusion limits on the ALP parameters, along with the current
        bound (hatched), in the \(\xi^2\alpha^{(2)}_{K^+\pi^0}\) --
        \(\xi^2\beta^{(2)}_{K^+\pi^0}\) plane. We obtain this by a
        speculative reduction of the  experimental and the theoretical
        uncertainties by 50\%. 
    }
	\label{fig:alpha-beta-contour-future}
\end{figure}

Note that, fitting the NP signal against the residual fluctuations
around the experimental best-fit would result in much stronger
constraints on \(\xi^2 \alpha^{(2)}_{K^+\pi^0}\) and \(\xi^2
\beta^{(2)}_{K^+\pi^0}\).  However, this approach assumes the true SM
value to coincide with the experimental best-fit point, and we do not
take this route.  Instead, we estimate the theoretical SM spectrum by
taking into account the full error in the FF computations, which
dominates over the experimental precision.  Naturally, these constraints
would become much stronger with reduced error in the FF computation,
assuming the SM prediction gets closer to the experimental number, with
the other possibility pointing to a discovery.  In this spirit, we
perform a simplistic analysis to estimate the reach of future
experiments by reducing the experimental error and assuming more precise
theoretical predictions. We keep the central values of the experimental
measurements and the theoretical predictions same as the current values,
and then reduce both the experimental and theoretical uncertainties by a
factor of 2.  In \Cref{fig:alpha-beta-contour-future}, we show the
\(95\%\) C.L.  allowed region for \(\xi^2 \alpha^{(2)}_{K^+\pi^0}\) and
\(\xi^2 \beta^{(2)}_{K^+\pi^0}\) as obtained from that analysis.

Before concluding this section, note that, we compute the MC event
distribution, corresponding to the NP signal, from our analytical
calculations as follows. Using our analytical computation of the
SM distribution and the SM MC Dalitz distribution (given by the NA48/2
collaboration) we can compute a bin-by-bin (as a function of
\((E_\pi,E_\ell)\)) scaling factor, let's call it
\(\mathcal{R}(E_\pi,E\mu)\), which represents the difference between the
analytical and the simulated results. We then multiply this function
with the analytical form of the NP distribution to get our proxy for the
MC simulated NP distribution,
\begin{align}
    \label{eq:proxy}
    \frac{d\Gamma}{dE_\pi dE_\mu}(E_\pi,E_\mu)\Huge\vert_\mathrm{final}
            &\equiv \mathcal{R}(E_\pi,E\mu) \otimes 
             \frac{d\Gamma}{dE_\pi
             dE_\mu}(E_\pi,E_\mu),
\end{align}
with the \(\otimes\) symbolically representing the bin-by-bin
multiplication.  Although this, somewhat pedestrian, approach is able to
capture all the essential effects sourced by the ALP, a more
sophisticated MC simulation might be of interest. 

\section{
    Direct vis \'a vis indirect detection
    \label{sec:ida}
}%
In the last section, we exclusively concentrated on the alterations of
the \(K^+_{\ell_3}\) theoretical expectations in the presence of an ALP.
This was motivated by our focus on indirect signatures of the ALP. It
is, however, instructive to look at the decay amplitudes for processes
involving the ALP itself (direct signatures) to fully appreciate the
benefits of the indirect stratagem.  This exercise also allows us to
point out the contributions to the ALP CC amplitudes from the
\(\mathcal{O}_W\) operator. 

Following the same exercise that led us to the \(K^+_{\ell_3}\)
Lagrangian in the A\(\chi\)PT (\cref{eq:finamp}), we can write down the
relevant interaction Lagrangian for \(K^+\to a\ell^+\nu\) as:
\begin{align}
    \label{eq:alpamp}
    \begin{split}
        \mathcal{L}_{a\ell^+\nu}\ \supset\ &i G_F V_{\bar{s}u} \:  
        \xi\Big[ \left(\alpha^{(1)}_{K^+ a}   
            + i \tilde{\alpha}^{(1)}_{K^+ a}\right)
            \left(
                K^{+} \partial_\mu \hat{a} - \partial_\mu K^+ \hat{a}
            \right) 
            +\left(\beta^{(1)}_{K^+ a} + i \tilde{\beta}^{(1)}_{K^+ a}   \right)
              \partial_\mu\left(K^{+}  \hat{a}\right)\Big]
    j^\mu_{-,\,\ell}\,,\\
    \quad\text{with}\quad
    & \alpha^{(1)}_{K^+a}\ =\ -\frac{1}{2}\left(
    C_{LR}^3-C_R^3 + \sqrt{3} (C_{LR}^8-C_R^8)\right),\;
    \\
    &\beta^{(1)}_{K^+ a}\ =\ -\frac{\sqrt{3}}{2}C_L^8,\quad
    \tilde{\beta}^{(1)}_{K^+ a}\ =\ -\tilde{\alpha}^{(1)}_{K^+ a}\ =\ C_W.
    \end{split}
\end{align} 
In the limit that the final state lepton mass goes to zero (for
simplicity), the corresponding amplitude-squared is proportional to:
\begin{align}
    \label{eq:Ka3_ampsq}
    |\mathcal{A}|^2_{K^+\to a\ell^+\nu}\propto\ 
        \xi^2\left(\left|\alpha^{(1)}_{K^+a}\right|^2 +
        \left|\tilde{\alpha}^{(1)}_{K^+a}\right|^2\right)\propto 
        \xi^2\left[\left(
            C_{LR}^3-C_R^3 + \sqrt{3} (C_{LR}^8-C_R^8)
        \right)^2 + (2C_W)^2\right].
\end{align}
Comments are in order about the strength of the \(K^+\to a\ell^+\nu\)
amplitude-squared. The first noteworthy thing is that the amplitude with
the ALP in the final state is of the order \(\xi^2\), the same as the
deviation of the corresponding pion amplitudes from the SM expectations.
The second thing to point out is that the amplitude gets a contribution
from the new \(C_W\) coefficient, which---being purely imaginary---does
not interfere with the other contributions. The consequence of this is
that there exists no limit where the different Wilson coefficients
conspire to set the amplitude to zero unless \(C_W\) is itself zero.
This observation is relevant in the context of the so-called pion-phobia
\cite{Georgi:1986df, Krauss:1987ud} that is popular in the literature.
The condition for said pion-phobia is generally presented in terms of
the quark masses and the couplings of the ALP to the scalar
quark-currents. This definition is, of course, model dependent. In a
more model-independent way, pion-phobia is just the limit
\cite{Georgi:1986df} where the \(K^+(\pi^+)\to a\ell^+\nu\) amplitude
vanishes, signifying, in effect, a flat direction in the plane of Wilson
coefficients.  It has, correctly, been pointed out that said flat
direction is not stable under renormalization \cite{Bauer:2020jbp}.
However, \cref{eq:Ka3_ampsq} shows that even at tree level this
cancellation does not exist when \(\mathcal{O}_W\) is taken into account. 

If we assume \(C_W=0\), there does exist a limit where the amplitude
goes to zero, i.e, \(\alpha^{(1)}_{K^+a}=0\). However, it is important
to realize that even in this limit, the effects of the ALP does not
decouple from the observables corresponding to the CC Lagrangian of the
mesons. To see this, we recall the \(K^+_{\ell_3}\) Lagrangian
(\cref{eq:finamp}) from the last section. As we can clearly see, by
comparing \cref{eq:albetDef} and \cref{eq:alpamp}, the effective
coefficient governing the deviation of the \(K^+_{\ell_3}\) amplitude
from SM expectations (viz.  \(\alpha^{(2)}_{K^+\pi^0}\)) expressed in
terms of \(\alpha^{(1)}_{K^+a}\),
\begin{equation}
    \label{eq:abK}
    \alpha^{(2)}_{K^+\pi^0}= \frac{C_3}{2}\left(\alpha^{(1)}_{K^+a}
    +\frac{C_3}{4}\right)\,,
\end{equation} 
remains non-zero even in the limit where \(\alpha^{(1)}_{K^+a}\) is
taken to be zero.  Consequently, even in the limit the \(K^+\to
a\ell^+\nu\) amplitude is suppressed, the NP contribution to the
\(K^+_{\ell_3}\) amplitude continues to be sizeable. This implies, even
in the pion-phobic limit the signatures of the ALP are not `invisible',
they are just buried in the distribution data of the processes with SM
final states.  Therefore, the condition for true pion-phobia is actually
more non-trivial. For it to happen, both the \(K^\pm\to a\ell\nu\)
amplitude and the deviation to the \(K^+_{\ell_3}\) amplitude
simultaneously need to  go to zero. Therefore, in terms of the Wilson
coefficients, the condition for `pion-phobia' is: 
\begin{align}
    \label{eq:PP_gen}
    C_W&\ =\ 0,\quad C_{LR}^3\ =\ C_R^3,\;\;\mathrm{and}\;\;\,C_{LR}^8=C_R^8,
\end{align}
at leading order in \(\xi^2\).  This is the limit where the ALP coupling
to the vectorial (RH) quark-currents identically cancels out the ALP
couplings to the scalar quark-currents.  It is easy to see that in this
limit, even the \(\pi^+\to a\ell\nu\) amplitude and the corresponding NP
effects to the \(\pi^+\to \pi^0 \ell\nu\) amplitude also go to zero.  To
explicitly see this, note that the \(\pi^+\to a\ell\nu\)
amplitude-squared is proportional to:
\begin{align}
    \label{eq:pi3_ampsq}
    \left|\alpha^{(1)}_{\pi^+a}\right|^2\ \propto\ \frac{(C_3)^2}{4},
\end{align}
in the limit \(C_W=0\). The  corresponding NP contribution to
\(\pi^+\to \pi^0 \ell\nu\) in terms of this \(\alpha^{(1)}_{\pi^+a}\)
is:
\begin{align}
    \label{eq:abP}
    &\left|\alpha^{(2)}_{\pi^+\pi^0}\right|^2\ \propto\  
    \frac{(C_3)^2}{4}\left(
            \alpha^{(1)}_{\pi^+a}+\frac{C_3}{4}
        \right)^2.
\end{align}
Clearly, in the limit given in \cref{eq:PP_gen}, this factor goes to
zero.  However, as we can see, the \(\pi^+\) case is slightly different
from the \(K^+\) decay case. We see that for the \(\pi^+\), a vanishing
of the \(\pi^+\!\to a\ell\nu\) amplitude-squared (i.e. $C_3=0, C_W=0$)
implies the vanishing of NP effects in the \(\pi^+\!\to \pi^0\ell\nu\)
amplitude-squared as well.  This is because unlike the \(t_L^8\) current
that contributes in \(K^+\) decay, the \(t_L^3\) current that
contributes to the \(\pi^+\) decay is EW vev suppressed and does not
contribute at this order. Hence, the difference is a consequence of
the global \(SU(2)_L\) symmetry that we discussed in 
\cref{sec:lag}. 

\section{
    Sum rules in meson decays
    \label{sec:sr}
} 
In this section, we discuss a way of identifying the presence and the
nature of an ALP in the chiral Lagrangian. We do this by formulating 
sum rules involving the form factors corresponding to leptonic
amplitudes of the light,  viz. \(\pi^0,\eta,\pi^+\) and \(K^+\),
mesons. One of the sums we discuss reduces to unity in the SM limit,  but
deviates from unity in the presence of an ALP. We can use the sign of
this deviation to distinguish between a meson Lagrangian where the ALPs
enter through mixing alone from that where there are EW interactions of
the ALP in the flavor basis itself. The sum rule, in principle, is a way
to tackle the EFT `inverse problem' and gives us a handle on
the differentiation of UV scenarios.

In deriving the sum rules, we work in the limit \(m_\ell\to 0\) where
the operators corresponding to \(\beta^{(1,2)}\) vanish and it is only
the operators corresponding to \(\alpha^{(1,2)}\) that contribute.
Hence, we need only to concentrate on processes with the electron in the
final state. As discussed in detail in the last section, this results in
the vanishing of the momentum-dependent effect and the net modification
is an overall scaling of the matrix element. 

In the SM, owing to the completeness of the \(\pi^0\)-\(\eta\) basis
(neglecting mixing with \(\eta^\prime\)), we have:
\begin{equation}
\label{eq:sumSM}
    \frac{1}{4}\left|f_{+,\,\mathrm{SM}}^{K^{+}\pi^0}(0)\right|^2
    + \frac{3}{4}\left|f_{+,\,\mathrm{SM}}^{K^{+}\eta}(0)\right|^2=\ 1.
\end{equation}
Here, the pre-factors of the FFs are the corresponding group theory
factors. In the presence of the ALP, this relationship is obviously
modified. The complete basis now includes the ALP, and this is reflected
in the sum of the form factors.  In the A\(\chi\)PT,  owing to the
redefinition of the physical \(\pi^0\) and the \(\eta\) mesons and due
to the introduction of new operators, the FFs are modified, as shown in
\cref{eq:ampDef,eq:albetDef}. These effective FFs, as we discussed, are
the objects which are extracted by the experiments. Therefore, in the
A\(\chi\)PT, these effective FFs are the natural candidates for the
construction of the sum.  In terms of these FFs, we find after some
algebra:
\begin{equation}
    \label{eq:modsum} 
    \frac{1}{4} \left|\tilde{f}_{+}^{K^{+}\pi^0}(0)\right|^2 +
    \frac{3}{4} \left|\tilde{f}_{+}^{K^{+}\eta}(0)\right|^2 
    =\ 1 
        - \dfrac{\xi^2}{16}\left(C_{LR}^3 - C_R^3  + \sqrt{3}(C_{LR}^8 - C_R^8)\right)^2 
        + \xi^2\dfrac{3}{16}(C_L^8)^2
\end{equation}
We see that when \(C_L^8=0\), the sum above is identically less than
one, as expected from considerations of completeness. However, when
there are \(t_L^8\) breaking interaction between the ALP and the mesons,
i.e.  \(C_L^8\neq 0\), the sum can be greater than one. Therefore, a
positive deviation of the sum from unity is not only possible, it
uniquely signals a \(t^8_L\) breaking interaction between the ALP and
the quarks in the UV.  That is, this sum can not only tell us about the
presence of an ALP in the chiral Lagrangian, but it can also tell us
about the corresponding UV model. The latter would not have been
possible by just looking at the deviations from SM expectations of the
individual decay widths . 

Before concluding, it is instructive to look at the corresponding sum
for a \(\pi^+\) in place of a \(K^+\):
\begin{equation}
    \label{eq:sumvisPi}
    \left|\tilde{f}_{+}^{\pi^{+}\pi^0}(0)\right|^2
    + \left|\tilde{f}_{+}^{\pi^{+}\eta}(0)\right|^2 =\ 1 - \dfrac{\xi^2}{4}C_3^2.
\end{equation} 
As is obvious, unlike the previous case, this sum is always less than
one.  As seen from \cref{eq:modsum}, a value of the sum greater than one
is possible only when \(C_L^8\ne 0\). That is, similar to the phenomenon
discussed in the last section, this result is sourced by the \(t_L^8\)
interaction of the ALP. The \(t_L^8\) counterpart for the \(\pi^+\)
sector is the \(t_L^3\). Now, as stressed in the first section, any
\(t_L^3\) interaction of the ALP breaks the \(SU(2)_L\) subgroup of
\(SU(3)_L\), hence, must be electroweak vev suppressed. This particular
result, along with the other \(SU(3)_L\) breaking effects discussed
throughout the paper, vindicates our choice of working with the
\(K^\pm\) decays as opposed to \(\pi^\pm\) decays. 

\section{
    Conclusion
    \label{sec:con}
}
A low-lying ALP leaves its signatures in amplitudes corresponding to SM
processes, signatures that are manifest at the tree level itself. These
signatures will be seen as variations from SM expectations in
conventional observables of flavor physics, for example, form factors,
differential distributions, decay rates etc. Therefore, indirect
detection techniques, like the ones discussed in this work, open up
novel avenues to look for ALPs and are complementary to standard
direct detection searches.  Furthermore, the tree level modification to
SM physics behoves us to undertake a careful study of the ALP-meson
Lagrangian in the light of the existing flavor physics anomalies.  The
results presented in this work, whether the data-driven analysis
corresponding to the \(K^\pm\) decay or the sum rules constructed out of
the form factors, are proof-of-concept examples that can be generalized
and used in the context of other observables.  Needless to stress that
the efficacy of such indirect techniques will only increase with the
inevitable improvements in lattice computations and with more precise
measurements of SM observables.  We expect that the methods discussed in
this work will be further generalized and applied to constrain the ALP
parameter space by focussing on data sets to be obtained from the
plethora of ongoing and upcoming flavor physics experiments.

\vskip 0.5em
\acknowledgments
    The authors thank Dmitry Madigozhin for providing them with the
    source of the NA48/2 dataset. The research of SG is supported by the
    NSF grant PHY-2014165.

\bibliographystyle{JHEP}
\bibliography{Axions}

\providecommand{\href}[2]{#2}\begingroup\raggedright\begin{thebibliography}{10}

\bibitem{tHooft:1976rip}
G.~'t~Hooft, \emph{{Symmetry Breaking Through Bell-Jackiw Anomalies}},
  \href{https://doi.org/10.1103/PhysRevLett.37.8}{\emph{Phys. Rev. Lett.}
  {\bfseries 37} (1976) 8}.

\bibitem{Peccei:1977hh}
R.D.~Peccei and H.R.~Quinn, \emph{{CP Conservation in the Presence of
  Instantons}}, \href{https://doi.org/10.1103/PhysRevLett.38.1440}{\emph{Phys.
  Rev. Lett.} {\bfseries 38} (1977) 1440}.

\bibitem{Peccei:1977ur}
R.D.~Peccei and H.R.~Quinn, \emph{{Constraints Imposed by CP Conservation in
  the Presence of Instantons}},
  \href{https://doi.org/10.1103/PhysRevD.16.1791}{\emph{Phys. Rev. D}
  {\bfseries 16} (1977) 1791}.

\bibitem{Weinberg:1977ma}
S.~Weinberg, \emph{{A New Light Boson?}},
  \href{https://doi.org/10.1103/PhysRevLett.40.223}{\emph{Phys. Rev. Lett.}
  {\bfseries 40} (1978) 223}.

\bibitem{Wilczek:1977pj}
F.~Wilczek, \emph{{Problem of Strong $P$ and $T$ Invariance in the Presence of
  Instantons}}, \href{https://doi.org/10.1103/PhysRevLett.40.279}{\emph{Phys.
  Rev. Lett.} {\bfseries 40} (1978) 279}.

\bibitem{Preskill:1982cy}
J.~Preskill, M.B.~Wise and F.~Wilczek, \emph{{Cosmology of the Invisible
  Axion}}, \href{https://doi.org/10.1016/0370-2693(83)90637-8}{\emph{Phys.
  Lett. B} {\bfseries 120} (1983) 127}.

\bibitem{Abbott:1982af}
L.F.~Abbott and P.~Sikivie, \emph{{A Cosmological Bound on the Invisible
  Axion}}, \href{https://doi.org/10.1016/0370-2693(83)90638-X}{\emph{Phys.
  Lett. B} {\bfseries 120} (1983) 133}.

\bibitem{Dine:1982ah}
M.~Dine and W.~Fischler, \emph{{The Not So Harmless Axion}},
  \href{https://doi.org/10.1016/0370-2693(83)90639-1}{\emph{Phys. Lett. B}
  {\bfseries 120} (1983) 137}.

\bibitem{chikashige:1980ui}
Y.~Chikashige, R.N.~Mohapatra and R.D.~Peccei, \emph{{Are There Real Goldstone
  Bosons Associated with Broken Lepton Number?}},
  \href{https://doi.org/10.1016/0370-2693(81)90011-3}{\emph{Phys. Lett. B}
  {\bfseries 98} (1981) 265}.

\bibitem{Froggatt:1978nt}
C.D.~Froggatt and H.B.~Nielsen, \emph{{Hierarchy of Quark Masses, Cabibbo
  Angles and CP Violation}},
  \href{https://doi.org/10.1016/0550-3213(79)90316-X}{\emph{Nucl. Phys. B}
  {\bfseries 147} (1979) 277}.

\bibitem{Izawa:2002qk}
K.I.~Izawa, T.~Watari and T.~Yanagida, \emph{{Higher dimensional QCD without
  the strong CP problem}},
  \href{https://doi.org/10.1016/S0370-2693(02)01663-5}{\emph{Phys. Lett. B}
  {\bfseries 534} (2002) 93}
  [\href{https://arxiv.org/abs/hep-ph/0202171}{{\ttfamily hep-ph/0202171}}].

\bibitem{Witten:1984dg}
E.~Witten, \emph{{Some Properties of O(32) Superstrings}},
  \href{https://doi.org/10.1016/0370-2693(84)90422-2}{\emph{Phys. Lett. B}
  {\bfseries 149} (1984) 351}.

\bibitem{Svrcek:2006yi}
P.~Svrcek and E.~Witten, \emph{{Axions In String Theory}},
  \href{https://doi.org/10.1088/1126-6708/2006/06/051}{\emph{JHEP} {\bfseries
  06} (2006) 051} [\href{https://arxiv.org/abs/hep-th/0605206}{{\ttfamily
  hep-th/0605206}}].

\bibitem{Arvanitaki:2009fg}
A.~Arvanitaki, S.~Dimopoulos, S.~Dubovsky, N.~Kaloper and J.~March-Russell,
  \emph{{String Axiverse}},
  \href{https://doi.org/10.1103/PhysRevD.81.123530}{\emph{Phys. Rev. D}
  {\bfseries 81} (2010) 123530}
  [\href{https://arxiv.org/abs/0905.4720}{{\ttfamily 0905.4720}}].

\bibitem{Chang:2000ii}
D.~Chang, W.-F.~Chang, C.-H.~Chou and W.-Y.~Keung, \emph{{Large two loop
  contributions to g-2 from a generic pseudoscalar boson}},
  \href{https://doi.org/10.1103/PhysRevD.63.091301}{\emph{Phys. Rev. D}
  {\bfseries 63} (2001) 091301}
  [\href{https://arxiv.org/abs/hep-ph/0009292}{{\ttfamily hep-ph/0009292}}].

\bibitem{Graham:2015cka}
P.W.~Graham, D.E.~Kaplan and S.~Rajendran, \emph{{Cosmological Relaxation of
  the Electroweak Scale}},
  \href{https://doi.org/10.1103/PhysRevLett.115.221801}{\emph{Phys. Rev. Lett.}
  {\bfseries 115} (2015) 221801}
  [\href{https://arxiv.org/abs/1504.07551}{{\ttfamily 1504.07551}}].

\bibitem{Jeong:2018jqe}
K.S.~Jeong, T.H.~Jung and C.S.~Shin, \emph{{Adiabatic electroweak baryogenesis
  driven by an axionlike particle}},
  \href{https://doi.org/10.1103/PhysRevD.101.035009}{\emph{Phys. Rev. D}
  {\bfseries 101} (2020) 035009}
  [\href{https://arxiv.org/abs/1811.03294}{{\ttfamily 1811.03294}}].

\bibitem{Nomura:2008ru}
Y.~Nomura and J.~Thaler, \emph{{Dark Matter through the Axion Portal}},
  \href{https://doi.org/10.1103/PhysRevD.79.075008}{\emph{Phys. Rev. D}
  {\bfseries 79} (2009) 075008}
  [\href{https://arxiv.org/abs/0810.5397}{{\ttfamily 0810.5397}}].

\bibitem{Rubakov:1997vp}
V.A.~Rubakov, \emph{{Grand unification and heavy axion}},
  \href{https://doi.org/10.1134/1.567390}{\emph{JETP Lett.} {\bfseries 65}
  (1997) 621} [\href{https://arxiv.org/abs/hep-ph/9703409}{{\ttfamily
  hep-ph/9703409}}].

\bibitem{Hook:2014cda}
A.~Hook, \emph{{Anomalous solutions to the strong CP problem}},
  \href{https://doi.org/10.1103/PhysRevLett.114.141801}{\emph{Phys. Rev. Lett.}
  {\bfseries 114} (2015) 141801}.

\bibitem{Fukuda:2015ana}
H.~Fukuda, K.~Harigaya, M.~Ibe and T.T.~Yanagida, \emph{{Model of visible QCD
  axion}}, \href{https://doi.org/10.1103/PhysRevD.92.015021}{\emph{Phys. Rev.
  D} {\bfseries 92} (2015) 015021}
  [\href{https://arxiv.org/abs/1504.06084}{{\ttfamily 1504.06084}}].

\bibitem{Marques-Tavares:2018cwm}
G.~Marques-Tavares and M.~Teo, \emph{{Light axions with large hadronic
  couplings}}, \href{https://doi.org/10.1007/JHEP05(2018)180}{\emph{JHEP}
  {\bfseries 05} (2018) 180}
  [\href{https://arxiv.org/abs/1803.07575}{{\ttfamily 1803.07575}}].

\bibitem{Gherghetta:2020keg}
T.~Gherghetta, V.V.~Khoze, A.~Pomarol and Y.~Shirman, \emph{{The Axion Mass
  from 5D Small Instantons}},
  \href{https://doi.org/10.1007/JHEP03(2020)063}{\emph{JHEP} {\bfseries 03}
  (2020) 063} [\href{https://arxiv.org/abs/2001.05610}{{\ttfamily
  2001.05610}}].

\bibitem{Irastorza:2018dyq}
I.G.~Irastorza and J.~Redondo, \emph{{New experimental approaches in the search
  for axion-like particles}},
  \href{https://doi.org/10.1016/j.ppnp.2018.05.003}{\emph{Prog. Part. Nucl.
  Phys.} {\bfseries 102} (2018) 89}
  [\href{https://arxiv.org/abs/1801.08127}{{\ttfamily 1801.08127}}].

\bibitem{Irastorza:2021tdu}
I.G.~Irastorza, \emph{{An introduction to axions and their detection}},  in
  \emph{{Les Houches summer school on Dark Matter}}, 9, 2021
  [\href{https://arxiv.org/abs/2109.07376}{{\ttfamily 2109.07376}}].

\bibitem{Alves:2017avw}
D.S.M.~Alves and N.~Weiner, \emph{{A viable QCD axion in the MeV mass range}},
  \href{https://doi.org/10.1007/JHEP07(2018)092}{\emph{JHEP} {\bfseries 07}
  (2018) 092} [\href{https://arxiv.org/abs/1710.03764}{{\ttfamily
  1710.03764}}].

\bibitem{Marciano:2016yhf}
W.J.~Marciano, A.~Masiero, P.~Paradisi and M.~Passera, \emph{{Contributions of
  axionlike particles to lepton dipole moments}},
  \href{https://doi.org/10.1103/PhysRevD.94.115033}{\emph{Phys. Rev. D}
  {\bfseries 94} (2016) 115033}
  [\href{https://arxiv.org/abs/1607.01022}{{\ttfamily 1607.01022}}].

\bibitem{Jaeckel:2015jla}
J.~Jaeckel and M.~Spannowsky, \emph{{Probing MeV to 90 GeV axion-like particles
  with LEP and LHC}},
  \href{https://doi.org/10.1016/j.physletb.2015.12.037}{\emph{Phys. Lett. B}
  {\bfseries 753} (2016) 482}
  [\href{https://arxiv.org/abs/1509.00476}{{\ttfamily 1509.00476}}].

\bibitem{Dobrich:2015jyk}
B.~D\"obrich, J.~Jaeckel, F.~Kahlhoefer, A.~Ringwald and K.~Schmidt-Hoberg,
  \emph{{ALPtraum: ALP production in proton beam dump experiments}},
  \href{https://doi.org/10.1007/JHEP02(2016)018}{\emph{JHEP} {\bfseries 02}
  (2016) 018} [\href{https://arxiv.org/abs/1512.03069}{{\ttfamily
  1512.03069}}].

\bibitem{Knapen:2016moh}
S.~Knapen, T.~Lin, H.K.~Lou and T.~Melia, \emph{{Searching for Axionlike
  Particles with Ultraperipheral Heavy-Ion Collisions}},
  \href{https://doi.org/10.1103/PhysRevLett.118.171801}{\emph{Phys. Rev. Lett.}
  {\bfseries 118} (2017) 171801}
  [\href{https://arxiv.org/abs/1607.06083}{{\ttfamily 1607.06083}}].

\bibitem{Bauer:2017ris}
M.~Bauer, M.~Neubert and A.~Thamm, \emph{{Collider Probes of Axion-Like
  Particles}}, \href{https://doi.org/10.1007/JHEP12(2017)044}{\emph{JHEP}
  {\bfseries 12} (2017) 044}
  [\href{https://arxiv.org/abs/1708.00443}{{\ttfamily 1708.00443}}].

\bibitem{Bauer:2018uxu}
M.~Bauer, M.~Heiles, M.~Neubert and A.~Thamm, \emph{{Axion-Like Particles at
  Future Colliders}},
  \href{https://doi.org/10.1140/epjc/s10052-019-6587-9}{\emph{Eur. Phys. J. C}
  {\bfseries 79} (2019) 74} [\href{https://arxiv.org/abs/1808.10323}{{\ttfamily
  1808.10323}}].

\bibitem{CidVidal:2018blh}
X.~Cid~Vidal, A.~Mariotti, D.~Redigolo, F.~Sala and K.~Tobioka, \emph{{New
  Axion Searches at Flavor Factories}},
  \href{https://doi.org/10.1007/JHEP01(2019)113}{\emph{JHEP} {\bfseries 01}
  (2019) 113} [\href{https://arxiv.org/abs/1810.09452}{{\ttfamily
  1810.09452}}].

\bibitem{Aloni:2018vki}
D.~Aloni, Y.~Soreq and M.~Williams, \emph{{Coupling QCD-Scale Axionlike
  Particles to Gluons}},
  \href{https://doi.org/10.1103/PhysRevLett.123.031803}{\emph{Phys. Rev. Lett.}
  {\bfseries 123} (2019) 031803}
  [\href{https://arxiv.org/abs/1811.03474}{{\ttfamily 1811.03474}}].

\bibitem{Bauer:2021mvw}
M.~Bauer, M.~Neubert, S.~Renner, M.~Schnubel and A.~Thamm, \emph{{Flavor probes
  of axion-like particles}},
  \href{https://arxiv.org/abs/2110.10698}{{\ttfamily 2110.10698}}.

\bibitem{Chakraborty:2021wda}
S.~Chakraborty, M.~Kraus, V.~Loladze, T.~Okui and K.~Tobioka, \emph{{Heavy QCD
  axion in $b\to s$ transition: Enhanced limits and projections}},
  \href{https://doi.org/10.1103/PhysRevD.104.055036}{\emph{Phys. Rev. D}
  {\bfseries 104} (2021) 055036}
  [\href{https://arxiv.org/abs/2102.04474}{{\ttfamily 2102.04474}}].

\bibitem{Bertholet:2021hjl}
E.~Bertholet, S.~Chakraborty, V.~Loladze, T.~Okui, A.~Soffer and K.~Tobioka,
  \emph{{Heavy QCD Axion at Belle II: Displaced and Prompt Signals}},
  \href{https://arxiv.org/abs/2108.10331}{{\ttfamily 2108.10331}}.

\bibitem{Freytsis:2009ct}
M.~Freytsis, Z.~Ligeti and J.~Thaler, \emph{{Constraining the Axion Portal with
  $B \to K l^+ l^-$}},
  \href{https://doi.org/10.1103/PhysRevD.81.034001}{\emph{Phys. Rev. D}
  {\bfseries 81} (2010) 034001}
  [\href{https://arxiv.org/abs/0911.5355}{{\ttfamily 0911.5355}}].

\bibitem{Bjorkeroth:2018dzu}
F.~Bj\"orkeroth, E.J.~Chun and S.F.~King, \emph{{Flavourful Axion
  Phenomenology}}, \href{https://doi.org/10.1007/JHEP08(2018)117}{\emph{JHEP}
  {\bfseries 08} (2018) 117}.

\bibitem{Altmannshofer:2019yji}
W.~Altmannshofer, S.~Gori and D.J.~Robinson, \emph{{Constraining axionlike
  particles from rare pion decays}},
  \href{https://doi.org/10.1103/PhysRevD.101.075002}{\emph{Phys. Rev. D}
  {\bfseries 101} (2020) 075002}
  [\href{https://arxiv.org/abs/1909.00005}{{\ttfamily 1909.00005}}].

\bibitem{Ishida:2020oxl}
H.~Ishida, S.~Matsuzaki and Y.~Shigekami, \emph{{New perspective in searching
  for axionlike particles from flavor physics}},
  \href{https://doi.org/10.1103/PhysRevD.103.095022}{\emph{Phys. Rev. D}
  {\bfseries 103} (2021) 095022}
  [\href{https://arxiv.org/abs/2006.02725}{{\ttfamily 2006.02725}}].

\bibitem{PIENU:2021clt}
{\scshape PIENU} collaboration, \emph{{Search for three body pion decays
  ${\pi}^+{\to}l^+{\nu}X$}},
  \href{https://doi.org/10.1103/PhysRevD.103.052006}{\emph{Phys. Rev. D}
  {\bfseries 103} (2021) 052006}
  [\href{https://arxiv.org/abs/2101.07381}{{\ttfamily 2101.07381}}].

\bibitem{Georgi:1986df}
H.~Georgi, D.B.~Kaplan and L.~Randall, \emph{{Manifesting the Invisible Axion
  at Low-energies}},
  \href{https://doi.org/10.1016/0370-2693(86)90688-X}{\emph{Phys. Lett.}
  {\bfseries 169B} (1986) 73}.

\bibitem{Bauer:2020jbp}
M.~Bauer, M.~Neubert, S.~Renner, M.~Schnubel and A.~Thamm, \emph{{The
  Low-Energy Effective Theory of Axions and ALPs}},
  \href{https://doi.org/10.1007/JHEP04(2021)063}{\emph{JHEP} {\bfseries 04}
  (2021) 063} [\href{https://arxiv.org/abs/2012.12272}{{\ttfamily
  2012.12272}}].

\bibitem{Bauer:2021wjo}
M.~Bauer, M.~Neubert, S.~Renner, M.~Schnubel and A.~Thamm, \emph{{Consistent
  Treatment of Axions in the Weak Chiral Lagrangian}},
  \href{https://doi.org/10.1103/PhysRevLett.127.081803}{\emph{Phys. Rev. Lett.}
  {\bfseries 127} (2021) 081803}
  [\href{https://arxiv.org/abs/2102.13112}{{\ttfamily 2102.13112}}].

\bibitem{Patt:2006fw}
B.~Patt and F.~Wilczek, \emph{{Higgs-field portal into hidden sectors}},
  \href{https://arxiv.org/abs/hep-ph/0605188}{{\ttfamily hep-ph/0605188}}.

\bibitem{Gell-Mann:1962yej}
M.~Gell-Mann, \emph{{Symmetries of baryons and mesons}},
  \href{https://doi.org/10.1103/PhysRev.125.1067}{\emph{Phys. Rev.} {\bfseries
  125} (1962) 1067}.

\bibitem{Okubo1}
S.~Okubo, \emph{{Note on Unitary Symmetry in Strong Interactions}},
  \href{https://doi.org/10.1143/PTP.27.949}{\emph{Progress of Theoretical
  Physics} {\bfseries 27} (1962) 949}.

\bibitem{Leutwyler:1993iq}
H.~Leutwyler, \emph{{On the foundations of chiral perturbation theory}},
  \href{https://doi.org/10.1006/aphy.1994.1094}{\emph{Annals Phys.} {\bfseries
  235} (1994) 165} [\href{https://arxiv.org/abs/hep-ph/9311274}{{\ttfamily
  hep-ph/9311274}}].

\bibitem{Lazzeroni:2018glh}
{\scshape NA48/2} collaboration, \emph{{Measurement of the form factors of
  charged kaon semileptonic decays}},
  \href{https://doi.org/10.1007/JHEP10(2018)150}{\emph{JHEP} {\bfseries 10}
  (2018) 150} [\href{https://arxiv.org/abs/1808.09041}{{\ttfamily
  1808.09041}}].

\bibitem{Zyla:2020zbs}
{\scshape Particle Data Group} collaboration, \emph{{Review of Particle
  Physics}}, \href{https://doi.org/10.1093/ptep/ptaa104}{\emph{PTEP} {\bfseries
  2020} (2020) 083C01}.

\bibitem{Carrasco:2016kpy}
N.~Carrasco, P.~Lami, V.~Lubicz, L.~Riggio, S.~Simula and C.~Tarantino,
  \emph{{$K \to \pi$ semileptonic form factors with $N_f=2+1+1$ twisted mass
  fermions}}, \href{https://doi.org/10.1103/PhysRevD.93.114512}{\emph{Phys.
  Rev. D} {\bfseries 93} (2016) 114512}
  [\href{https://arxiv.org/abs/1602.04113}{{\ttfamily 1602.04113}}].

\bibitem{Cheng:2021kjg}
H.-C.~Cheng, L.~Li and E.~Salvioni, \emph{{A Theory of Dark Pions}},
  \href{https://arxiv.org/abs/2110.10691}{{\ttfamily 2110.10691}}.

\bibitem{Callan:1969sn}
J.~Callan, Curtis~G., S.R.~Coleman, J.~Wess and B.~Zumino, \emph{{Structure of
  phenomenological Lagrangians. 2.}},
  \href{https://doi.org/10.1103/PhysRev.177.2247}{\emph{Phys. Rev.} {\bfseries
  177} (1969) 2247}.

\bibitem{Kaplan:2005es}
D.B.~Kaplan, \emph{{Five lectures on effective field theory}},  10, 2005
  [\href{https://arxiv.org/abs/nucl-th/0510023}{{\ttfamily nucl-th/0510023}}].

\bibitem{Donoghue:1996zn}
J.F.~Donoghue and A.F.~Perez, \emph{{The Electromagnetic mass differences of
  pions and kaons}},
  \href{https://doi.org/10.1103/PhysRevD.55.7075}{\emph{Phys. Rev. D}
  {\bfseries 55} (1997) 7075}
  [\href{https://arxiv.org/abs/hep-ph/9611331}{{\ttfamily hep-ph/9611331}}].

\bibitem{Christ:2010dd}
N.H.~Christ, C.~Dawson, T.~Izubuchi, C.~Jung, Q.~Liu, R.D.~Mawhinney et~al.,
  \emph{{The $\eta$ and $\eta^\prime$ mesons from Lattice QCD}},
  \href{https://doi.org/10.1103/PhysRevLett.105.241601}{\emph{Phys. Rev. Lett.}
  {\bfseries 105} (2010) 241601}
  [\href{https://arxiv.org/abs/1002.2999}{{\ttfamily 1002.2999}}].

\bibitem{Bali:2021vsa}
{\scshape RQCD} collaboration, \emph{{Properties of the $\eta$ and
  $\eta^{\prime}$ mesons: Masses, decay constants and gluonic matrix
  elements}}, {\emph{PoS} {\bfseries LATTICE2021} (2021) 286}
  [\href{https://arxiv.org/abs/2111.05656}{{\ttfamily 2111.05656}}].

\bibitem{Aoki:2013ldr}
S.~Aoki et~al., \emph{{Review of Lattice Results Concerning Low-Energy Particle
  Physics}}, \href{https://doi.org/10.1140/epjc/s10052-014-2890-7}{\emph{Eur.
  Phys. J. C} {\bfseries 74} (2014) 2890}
  [\href{https://arxiv.org/abs/1310.8555}{{\ttfamily 1310.8555}}].

\bibitem{Cirigliano:2001mk}
V.~Cirigliano, M.~Knecht, H.~Neufeld, H.~Rupertsberger and P.~Talavera,
  \emph{{Radiative corrections to K(l3) decays}},
  \href{https://doi.org/10.1007/s100520100825}{\emph{Eur. Phys. J. C}
  {\bfseries 23} (2002) 121}
  [\href{https://arxiv.org/abs/hep-ph/0110153}{{\ttfamily hep-ph/0110153}}].

\bibitem{Sirlin:1981ie}
A.~Sirlin, \emph{{Large m(W), m(Z) Behavior of the O(alpha) Corrections to
  Semileptonic Processes Mediated by W}},
  \href{https://doi.org/10.1016/0550-3213(82)90303-0}{\emph{Nucl. Phys. B}
  {\bfseries 196} (1982) 83}.

\bibitem{FlaviaNet:2008hpm}
{\scshape FlaviaNet Working Group on Kaon Decays} collaboration,
  \emph{{Precision tests of the Standard Model with leptonic and semileptonic
  kaon decays}},  in \emph{{5th International Workshop on e+ e- Collisions from
  Phi to Psi}}, 1, 2008 [\href{https://arxiv.org/abs/0801.1817}{{\ttfamily
  0801.1817}}].

\bibitem{Moulson:2017ive}
M.~Moulson, \emph{{Experimental determination of $V_{us}$ from kaon decays}},
  \href{https://doi.org/10.22323/1.291.0033}{\emph{PoS} {\bfseries CKM2016}
  (2017) 033} [\href{https://arxiv.org/abs/1704.04104}{{\ttfamily
  1704.04104}}].

\bibitem{Bernard:2009zm}
V.~Bernard, M.~Oertel, E.~Passemar and J.~Stern, \emph{{Dispersive
  representation and shape of the K(l3) form factors: Robustness}},
  \href{https://doi.org/10.1103/PhysRevD.80.034034}{\emph{Phys. Rev. D}
  {\bfseries 80} (2009) 034034}
  [\href{https://arxiv.org/abs/0903.1654}{{\ttfamily 0903.1654}}].

\bibitem{Marciano:2004uf}
W.J.~Marciano, \emph{{Precise determination of |V(us)| from lattice
  calculations of pseudoscalar decay constants}},
  \href{https://doi.org/10.1103/PhysRevLett.93.231803}{\emph{Phys. Rev. Lett.}
  {\bfseries 93} (2004) 231803}
  [\href{https://arxiv.org/abs/hep-ph/0402299}{{\ttfamily hep-ph/0402299}}].

\bibitem{zenodona4818}
{\scshape NA48/2} collaboration, M.~Dmitry and S.~Sergey, \emph{{NA48/2 program
  and data for calculation of charged kaon semileptonic form factors}},  Dec.,
  2019.
\newblock 10.5281/zenodo.3560600.

\bibitem{Krauss:1987ud}
L.M.~Krauss and D.J.~Nash, \emph{{A viable weak interaction axion?}},
  \href{https://doi.org/10.1016/0370-2693(88)91864-3}{\emph{Phys. Lett. B}
  {\bfseries 202} (1988) 560}.

\end{thebibliography}\endgroup

\end{document}